\documentclass[aip,jcp]{revtex4-1}

\usepackage[mathlines]{lineno} % Enable numbering of text and display math - needed for email
\usepackage{graphicx} % Include figure files - needed
\usepackage{amsmath} % more available functions for math mode - needed
\usepackage{geometry}                		% See geometry.pdf to learn the layout options. There are lots.
\usepackage{amssymb,color}
\usepackage{setspace}
\usepackage{footmisc}

\usepackage{bm}
\usepackage{subfigure}
\usepackage{amsmath}
\usepackage{xcolor}
\usepackage{xparse,xcoffins}

\ExplSyntaxOn
\NewCoffin\imagecoffin
\NewCoffin\labelcoffin

\keys_define:nn { miguel/label }
 {
  label   .tl_set:N = \l_miguel_label_tl,
  labelbox .bool_set:N = \l_miguel_label_box_bool,
  labelbox .default:n = true,
  fontsize .tl_set:N = \l_miguel_label_size_tl,
  fontsize .initial:n = \footnotesize,
  pos .choice:,
  pos/nw .code:n = \tl_set:Nn \l_miguel_label_pos_tl { left,up },
  pos/ne .code:n = \tl_set:Nn \l_miguel_label_pos_tl { right,up },
  pos/sw .code:n = \tl_set:Nn \l_miguel_label_pos_tl { left,down },
  pos/se .code:n = \tl_set:Nn \l_miguel_label_pos_tl { right,down },
  pos/n .code:n = \tl_set:Nn \l_miguel_label_pos_tl { hc,up },
  pos/w .code:n = \tl_set:Nn \l_miguel_label_pos_tl { left,vc },
  pos/s .code:n = \tl_set:Nn \l_miguel_label_pos_tl { hc,down },
  pos/e .code:n = \tl_set:Nn \l_miguel_label_pos_tl { right,vc },
  pos .initial:n = nw,
  unknown .code:n   = \clist_put_right:Nx \l_miguel_label_clist
                       { \l_keys_key_tl = \exp_not:n { #1 } }
 }
\clist_new:N \l_miguel_label_clist
\box_new:N \l_miguel_label_box
\box_new:N \l_miguel_label_image_box

\NewDocumentCommand{\xincludegraphics}{O{}m}
 {
  \group_begin:
  \tl_clear:N \l_miguel_label_tl
  \clist_clear:N \l_miguel_label_clist
  \keys_set:nn { miguel/label } { #1 }
  \tl_if_empty:NTF \l_miguel_label_tl
   {
    \miguel_includegraphics:Vn \l_miguel_label_clist { #2 }
   }
   {
    \SetHorizontalCoffin\imagecoffin
     {
      \miguel_includegraphics:Vn \l_miguel_label_clist { #2 }
     }
    \SetHorizontalCoffin\labelcoffin
     {
      \raisebox{\depth}
       {
        \bool_if:NTF \l_miguel_label_box_bool
         { \fcolorbox{white}{white}{\l_miguel_label_size_tl\l_miguel_label_tl} }
         { \l_miguel_label_size_tl\l_miguel_label_tl }
       }
     }
    \SetVerticalPole\imagecoffin{left}{3pt+\CoffinWidth\labelcoffin/2}
    \SetVerticalPole\imagecoffin{right}{\Width-3pt-\CoffinWidth\labelcoffin/2}
    \SetHorizontalPole\imagecoffin{up}{\Height-3pt-\CoffinHeight\labelcoffin/2}
    \SetHorizontalPole\imagecoffin{down}{3pt+\CoffinHeight\labelcoffin/2}
    \use:x{\JoinCoffins\imagecoffin[\l_miguel_label_pos_tl]\labelcoffin[vc,hc]} 
    \TypesetCoffin\imagecoffin
   }
   \group_end:
 }
\NewDocumentCommand{\setlabel}{m}
 {
  \keys_set:nn { miguel/label } { #1 }
 }

\cs_new_protected:Nn \miguel_includegraphics:nn
 {
  \includegraphics[#1]{#2}
 }
\cs_generate_variant:Nn \miguel_includegraphics:nn { V }

\ExplSyntaxOff

\makeatletter
\def\frontmatter@thefootnote{%
 \altaffilletter@sw{\@fnsymbol}{\@fnsymbol}{\csname c@\@mpfn\endcsname}%
}
\makeatother

\begin{document}

\title{Decoupling of single-particle and collective dynamics in arrested phase-separating glassy mixtures} 

\author{Konstantin N.~Moser}
\email{konstantin-moser@univie.ac.at}
\author{Christos N.~Likos}
\email{christos.likos@univie.ac.at}
\affiliation{Faculty of Physics, University of Vienna, Boltzmanngasse 5, 1090 Vienna, Austria}
\author{Vittoria Sposini}
\email{vittoria.sposini@unipd.it -- Corresponding author}
\affiliation{Faculty of Physics, University of Vienna, Boltzmanngasse 5, 1090 Vienna, Austria}
\affiliation{Dipartimento di Fisica e Astronomia ‘G.~Galilei’-- DFA, Sezione INFN, Universit{\`a} di Padova, Via Marzolo 8, 35131 Padova, Italy}

\date{\today}

\begin{abstract}
{\bf Abstract.} We investigate the structure and dynamics of a hard colloid--star polymer mixture in the range of its arrested phase separation, where an incipient demixing transition is interfering with a nearby vitrification line, 
focusing on the protein limit (smaller hard component). 
Soft-hard mixtures present a rich dynamics, influenced by different parameters such as the concentration of the soft and hard components, the softness of the potential, and the size ratio between the two components.
Using coarse-grained molecular dynamics simulations, we characterize the 
single-particle and collective dynamics of the hard colloidal tracers in the soft glassy matrix.
The hard tracers show diffusive behavior of the mean squared displacement accompanied by non-exponential relaxation of the intermediate scattering functions at intermediate length scales and non-Gaussian displacement distributions. 
Moreover, we show that the system exhibits arrested phase separation, leading to population splitting and decoupling between self- and collective dynamics of the hard colloids. 
Overall, we demonstrate that the interplay between arrested phase separation and glassiness leads to complex, multiscale phenomena that strongly influence the dynamics of the hard additives in the arrested matrix formed by the soft colloids. 
\end{abstract}

\maketitle

\section{Introduction}
\label{sec:intr}

The study of dynamic properties in soft matter systems displaying dynamical arrest is a topic investigated intensively by the scientific community, as it plays a crucial role in understanding and steering the rheological properties of materials from both the fundamental and the technological points of view. A very useful experimental approach to probe the microstructure and dynamics of a glassy soft material is microrheology, i.e., the introduction of small tracer particles whose dynamics is monitored by, e.g., optical techniques.~\cite{microrheology:book}
The motion of colloidal tracers in heterogeneous environments often exhibits an effective diffusive behavior that is distinct from the classical Brownian motion, and has received a lot of attention.
Examples in this direction cover nanoparticles diffusing in polymer solutions,~\cite{hofling2013anomalous,xue2020diffusion} in porous media,~\cite{kurzidim2009single, kim2009slow, spanner2011anomalous, ghosh2015non, sorichetti2023structure, edimeh2025dynamics} and in gels.~\cite{babu2008tracer} 
Glass-forming liquids are a paradigmatic class of systems exhibiting pronounced dynamic heterogeneity, in which both spatial and temporal fluctuations become increasingly prominent as the vitrification transition is approached.~\cite{berthier2011theoretical, arceri2022glasses} 
The study of colloidal tracer dynamics in such glassy environments thus provides valuable insight into the microscopic mechanisms governing relaxation and arrest in complex fluids.
Most previous studies have focused on glass-forming liquids characterized by hard-sphere-like repulsive interactions~\cite{sentjabrskaja2016anomalous, brizioli2022reciprocal,poling2019structure, edimeh2025dynamics} and/or attractive potential.~\cite{roberts2018tracer}
In contrast, the present work investigates tracer dynamics in a glassy matrix with a soft interparticle potential, a regime that remains comparatively less explored.
Soft potentials, obtained via specific coarse graining techniques, are used to describe effective interactions between macromolecules~\cite{likos2001effective, klapp2004effective} and differ fundamentally from hard-sphere potentials because they feature a finite or weakly divergent repulsive core.  
This property allows particles to interpenetrate to some extent, leading to markedly different structural and dynamic behavior.~\cite{likos2001effective} 
As a consequence, soft--hard mixtures also present a rich phase diagram due to multiple mechanisms of vitrification and melting. 
In addition, as their concentration increases, the inserted tracer particles cease to be mere probes of the underlying heterogeneous dynamics of the glassy matrix: instead, their presence modifies the structure of the system, and it can drive glass melting as well as (arrested) phase separation of the mixture. 
This less studied situation is the focus of the present study.

In this work, we focus in particular on the case of star polymers as soft component.~\cite{likos:prl:1998} 
Star polymers represent a valuable model system that interpolates between hard spheres and polymer chains: the control parameter is the number of arms (functionality) $f$, yielding hard-colloid behavior in the regime $f \gg 1$ and linear chains for $f = 1$ or $f = 2$. 
This property is reflected in the phase diagram of star polymers, which shows no crystallization for low $f$-values~\cite{watzlawek:prl:1999,rovigatti:acsnano:2014} and hard-sphere properties as $f \gg 1$, as also confirmed by experimental studies, in which crystallization has been shown to be favored by solvents of intermediate quality~\cite{stiakakis:pre:2010} and by the application of  shear.~\cite{ruizfranco:prl:2018}
As the formation of periodic crystals is hindered by kinetic effects as well as by polydispersity, a great deal of attention has been devoted to the vitrification of star polymers, i.e., the emergence of a structurally arrested, amorphous solid upon increasing the concentration for fixed $f$-values.~\cite{foffi:prl:2003,yang:epl:2010} 
Similarly to the case of crystallization, the glass formation of star polymers occurs only for sufficiently high values of the functionality, $f \geq f_{\rm c} \cong 35$.
For a range of values $f \gtrsim f_{\rm c}$, a glass can be formed by compression and remelt into an ergodic fluid upon further increase of the concentration, a property shared by glasses formed for systems interacting by a broad family of ultrasoft effective potentials.\cite{pamies:jcp:2009,berthier:pre:2010,ikeda:prl:2011, srivastava:prl:2013,miyazaki:prl:2016,miyazaki:jcp:2019,camerin:prx:2020,wuh:prr:2021,sposini:sm:2023,wuh:mm:2023}
The theoretical prediction on the glass transition of star polymers~\cite{foffi:prl:2003} has been experimentally confirmed in the work of Gupta {\it et al.},~\cite{gupta:nsc:2015} who employed a system of self-aggregating block copolymers as a proxy to star polymers. 
Moreover, experimentation with the same system has revealed that it satisfies the Stokes-Einstein relation all the way up to the glass transition,~\cite{gupta:prl:2015}, a property that can be attributed to the ultrasoft character of the interaction potential.
Additional work has been dedicated to the investigations of how the vitrification of star polymers can be affected by the presence of soft additives of size smaller than the stars.
Two characteristic examples are the addition of small linear homopolymer~\cite{stiakakis:prl:2002,lonetti:prl:2011} and of other star polymers of lower functionality.~\cite{zaccarelli:prl:2005,mayer:natmat:2008,mayer:mm:2009} 
In both cases, the effect of the additives on an arrested star polymer solution slightly above its vitrification concentration is to restore ergodicity, an effect caused by a weakening of the inter-star repulsions due to depletion effects of the smaller soft component. 
For the case of star-star mixtures, further increase of the additives' concentration leads to a variety of structurally arrested states.~\cite{mayer:natmat:2008,mayer:mm:2009}

\begin{figure}
    \centering
    \includegraphics[width=0.5\linewidth]{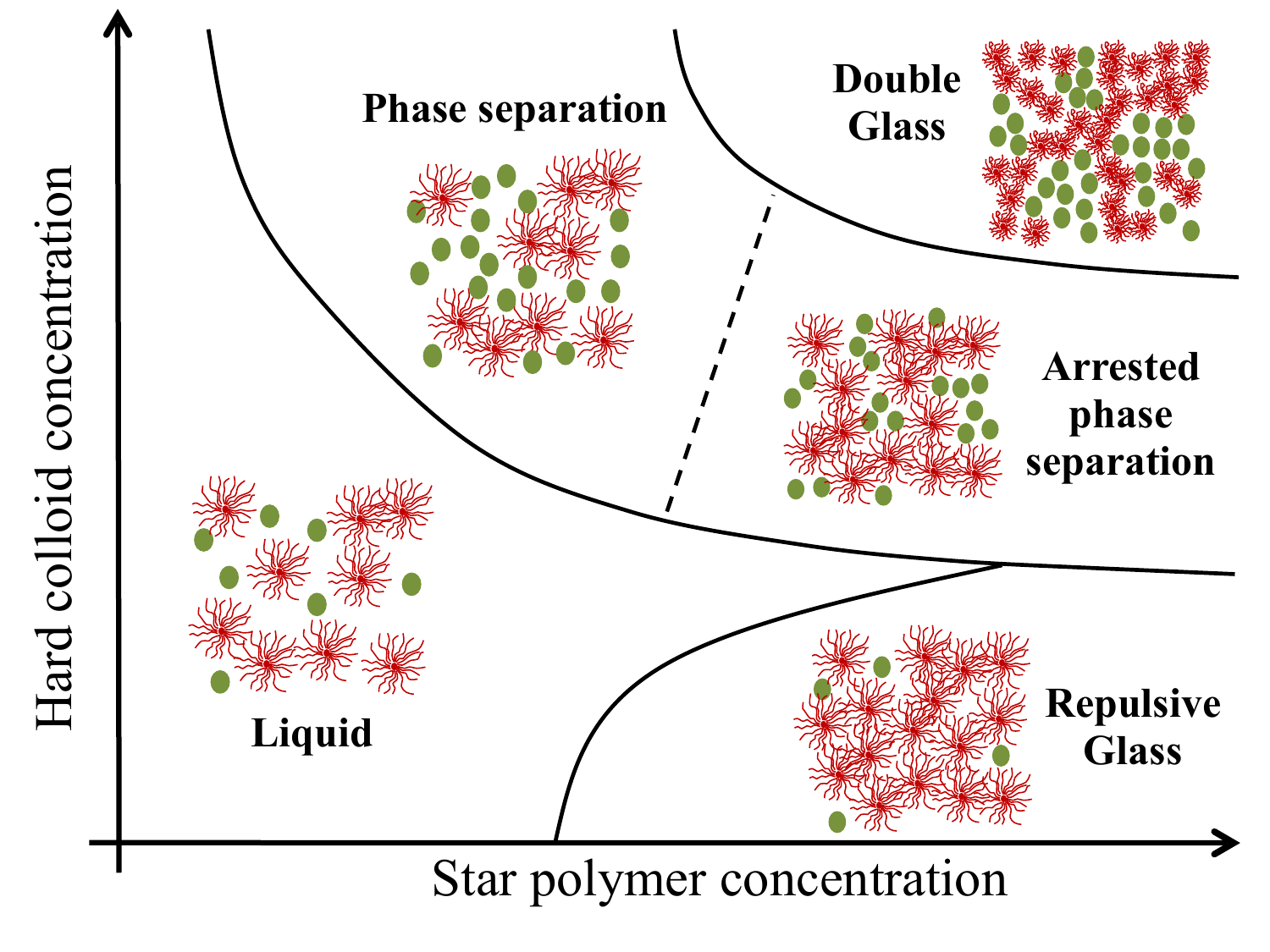}
    \caption{Schematic phase diagram  of a binary star polymer -- colloid mixture for high functionality stars and smaller colloids. 
    The horizontal and vertical axes represent the star polymer- and hard colloid concentrations, respectively. 
    Exact concentration values for which the various transitions occur depend on the softness of the potential and size ratio of the two components. 
    Redrawn with permission from Merola {\em et al.}~\cite{merola2018asymmetric}}
    \label{fig:phasediagr}
\end{figure}

Mixtures between hard colloids and soft star polymers have been examined in two complementary limiting cases. 
In the more conventional ``colloidal limit", the star polymers are smaller than the  colloids and thus they act as depletants, bringing about a demixing transition.~\cite{dzubiella:jcp:2002,wuq:acsami:2020} 
Recently, attention has been turned to the opposite, ``protein limit", in which the larger star polymers are the majority component and the added hard colloids act as modifiers of the glassy state formed by the soft colloidal stars.~\cite{truzzolillo:prl:2013,marzi:sm:2015,parisi:jcp:2021}
Explicit coarse-grained potentials derived in Ref.~\onlinecite{marzi2012coarse} have been used to explore in detail star polymer--hard colloid mixtures.
Fig.~\ref{fig:phasediagr} shows the typical phase diagram for this mixture based on experimental evidence and theoretical calculations.~\cite{marzi:sm:2015,merola2018asymmetric,parisi:jcp:2021}
At intermediate star polymer concentrations, where the pure star system approaches a glass transition, the addition of hard colloids drives a restoration of ergodicity from a repulsive glass to a liquid sate, followed thereafter by a dynamically arrested and inhomogeneous phase separated state that eventually ends in a double glass. 
Previous work has focused on a combination of rheological analysis, accompanied by Mode-Coupling Theory, to analyze the physical mechanisms leading to the glass melting and the dependence of the same on star functionality and size ratio.~\cite{truzzolillo:prl:2013,marzi:sm:2015}
A more detailed, microscopic analysis of the coupled structure and dynamics of the two components in the region surrounded by the soft glass and the incipient phase separation is, however, lacking.

The goal of this work is to explore and characterize the first two transitions (glass--liquid--phase separated) with a particular focus on the dynamics of the hard colloids as probes of the system’s heterogeneity. 
Using Molecular Dynamics simulations, we aim at elucidating how changes in the hard colloid concentration influence both structure and dynamics of the mixture. 
By analyzing the motion and relaxation behavior of the hard component, we study how the interplay between glassy dynamics in the soft matrix and incipient phase separation gives rise to complex, multi-scale phenomena.
The rest of the paper is organized as follows:
In section~\ref{sec:comp-model} we introduce our computational model and methods. 
Results and related discussions are presented in section~\ref{sec:res}, while in section~\ref{sec:concl} we report our conclusions.

\section{Computational model}
\label{sec:comp-model}
\subsection{Coarse-graining}

Both components, the hard colloids and the star polymers, are implemented in a coarse-grained fashion, using their centers as
effective coordinates, and employing thereby effective potentials $V_{ij}(r)$, where $r$ stands for the distance between the respective centers and $i,j \in \{H, S\}$, where $H$ ($S$) stands for the hard sphere (star polymer) component. 
Moreover, we introduce the hard sphere and star polymer sizes $\sigma_H$ and $\sigma_S$, respectively, whose precise meaning will be explained in what follows.

In previous, theoretical approaches,~\cite{truzzolillo:prl:2013,marzi:sm:2015,parisi:jcp:2021} colloids were modeled to interact with one another via a hard sphere potential of diameter $\sigma_H$. 
As this is rather inconvenient for the MD simulations we employ here, we replace it in this work by a very steep Weeks-Chandler-Anderson (WCA) potential:~\cite{weeks1971role}
\begin{equation}
    \beta V_{HH}(r) = 
    \begin{cases}
        4\epsilon\left[\left(\frac{\sigma_H}{r}\right)^{12} - \left(\frac{\sigma_H}{r}\right)^{6}\right] + \epsilon, 
        & r \leq  2^{1/6}\sigma_H,\\
        0, & r >  2^{1/6} \sigma_H.
    \end{cases}    
    \label{eq:Vhh}
\end{equation}
with $\beta=(k_{\rm B} T)^{-1}$, $T$ being the absolute temperature and $k_{\rm B}$ Boltzmann's constant; 
we set $\epsilon= 10^2$. 
The coarse grained potential for the star polymers is given by the expression:~\cite{Jusufi2001softhard, likos2001effective}
\begin{equation}
    \beta V_{SS}(r) = \frac{5}{18} f^{3/2}
    \begin{cases}
        - \ln \left(\dfrac{r}{\sigma_S}\right) + \dfrac{1}{1+\sqrt{f}/2}, & r<  \sigma_S,\\
        \dfrac{1}{1+\sqrt{f}/2}\left(\dfrac{\sigma_S}{r}\right) \exp\left( -\dfrac{\sqrt{f}}{2\sigma_S}(r-\sigma_S)\right), & r \ge  \sigma_S,
    \end{cases}    
    \label{eq:Vss}
\end{equation}
where $f$ is the functionality (number of arms of the star polymers) and $\sigma_S$ is the corona diameter, related to the star's radius of gyration $R_S^{\rm gyr}$ via~\cite{likos:prl:1998} $\sigma_S = 4\,R_S^{\rm gyr}/3$.

The cross interaction between soft and hard colloids was derived by Marzi {\em et al.}~\cite{marzi2012coarse} and reads as:
\begin{equation}
    \beta V_{SH}(r) = \frac{5}{18} f^{3/2}
    \begin{cases}
        \infty, & r<  R_H,\\
        -\int_r^\infty F_{SH}(r'-R_H) {\rm d}r', & r \ge  R_H,
    \end{cases}    
    \label{eq:Vsh}
\end{equation}
where $R_H=\sigma_H/2$ and the cross force $F_{SH}(z)$ is obtained by the following integral:
\begin{equation}
    F_{SH}(z)=\frac{\pi R_H}{(z+R_H)^2} \int_z^{s_\mathrm{max}} \left(z^2+2 R_H z +s^2\right)\left[\Pi(s)- \Pi(s+t)\right]{\rm d}s,
\end{equation}
where $z=r-R_H$, $s_\mathrm{max}=\sqrt{z(z+2R_H)}$, and $t=(s_\mathrm{max}^2-s^2)/s$. 
The osmotic pressure $\Pi(s)$ is expressed as
\begin{equation}
    \beta \Pi(s) = \frac{\Lambda }{s^2} f^{3/2}
    \begin{cases}
        s^{-1}, & s \le  R_S,\\
        \dfrac{1+2\kappa^2s^2}{1+2\kappa^2R_S^2} \exp\left( -\kappa^2(s^2-R_S^2)\right), & s >  R_S,
    \end{cases}    
    \label{eq:osm-press}
\end{equation}
where $R_S=\sigma_S/2$ and the parameters $\kappa$ and $\Lambda$ are related to the functionality $f$ and are fixed to $\kappa=0.96R_S$ and $\Lambda=5/36\pi$.~\cite{marzi2012coarse}
In line with the experimental systems and theoretical studies reported in Ref.~\onlinecite{parisi:jcp:2021}, we fix $f = 166$, and $\sigma_H=0.667\sigma_S$, corresponding to a size ratio $R_S^\mathrm{gyr}/R_H=2.25$. 
\begin{figure}
    \centering
    \includegraphics[width=0.5\linewidth]{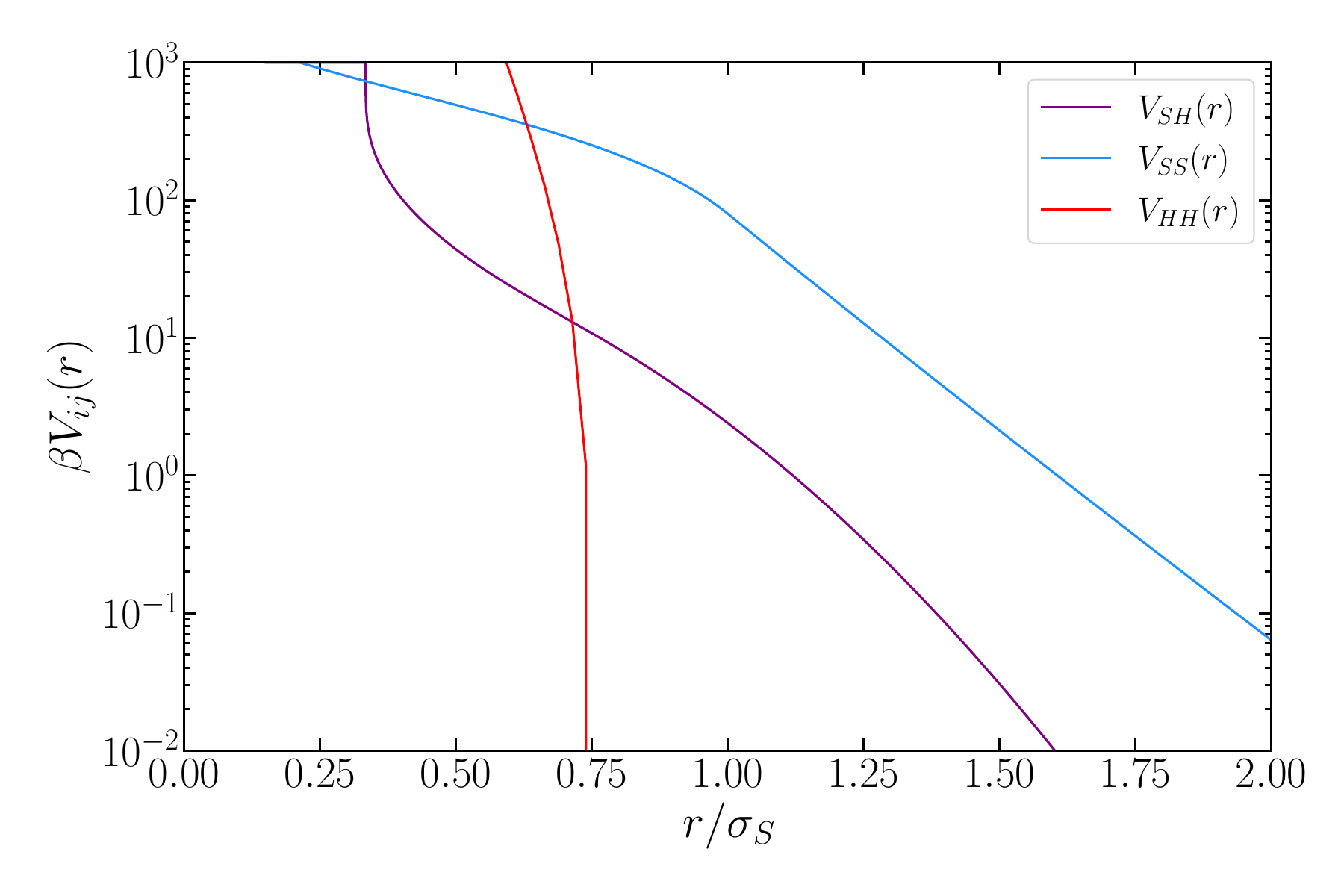}
    \caption{Coarse-grained potentials as used in the simulations with $f = 166$ and $\sigma_H = 0.667 \sigma_S$.}
    \label{fig:potentials}
\end{figure}
In Fig.~\ref{fig:potentials}, we show the coarse-grained potentials for these values of the parameters. 

\subsection{Molecular Dynamics simulations and system setup}
\label{sec:methods:simumlations} 

We run molecular dynamics simulations in LAMMPS~\cite{thompson2022lammps}. 
To sample the $NVT$-ensemble, we employ a Nose-Hoover thermostat and, for simplicity, we fix $T=1.0$ and $k_{\rm B} = 1$. 
In what follows, we also set the star corona diameter $\sigma_S = 1$ and assign to both the colloids and the stars the same mass $m = 1$, completing therefore the choice of units for mass, length and energy. 
Accordingly, the quantity $\tau = \sqrt{m\sigma_S^2/(k_{\rm B} T)}$ sets the unit of simulation time. 
The system is simulated in a cubic box $L \times L \times L$ with $L = 21.72\sigma_S$, having a volume $V = L^3 \cong 10\,246.6\,\sigma_S^3$.

Our goal is to examine a system of star polymers slightly above their vitrification line, so that the pure star system is glassy, and add different amounts of hard colloids, in accordance with experiment.~\cite{truzzolillo:prl:2013,marzi:sm:2015,parisi:jcp:2021}
We choose a star density $\rho_S\sigma_S^3 = 0.4$ for that purpose, which fixes the total number of stars in the box to $N_S = 4\,100$.
Creating a glassy state requires particular care: first of all, we have to suppress crystallization of the sample, as the perfect (fcc) crystal is the equilibrium structure;~\cite{watzlawek:prl:1999} and second, we need to quench the system slowly to obtain an equilibrated arrested state guaranteeing reproducibility of the results.

To avoid the periodic crystal, we introduce polydispersity in the diameters of the soft colloids $\sigma_S$ and obtain a glassy state instead.~\cite{berthier2023modern,Ninarello2017glassalgos,Pihlajamaa2023glasspolydisp} 
In particular, we split the $N_S$ total number of stars into three subpopulations, namely: $N_S^{<} = 2\,000$ stars with a smaller than average diameter $\sigma_S^{<} = 0.92\,\sigma_S$; 
$N_S^{\rm m} = 1\,200$ stars with diameter $\sigma_S^{\rm m} = \sigma_S$;
and $N_S^{>} = 900$ stars with a larger than average diameter $\sigma_S^{>} = 1.15\,\sigma_S$.
Evidently, $N_S^{<} + N_S^{\rm m} + N_S^{>} = N_S$; moreover, the mean of this distribution is 
$\langle \sigma_S \rangle \equiv (N_S^{<}\sigma_S^{<} + N_S^{\rm m} \sigma_S^{\rm m} + N_S^{>}\sigma_S^{>})/N_S = \sigma_S$, which is the value we use when evaluating the cross-interactions. 
In this way the polydspersity allows us to approach a glassy state for the stars but does not affect the cross-interactions.
Quenching to the glassy state requires additional care because, since the star-star potential is athermal, a standard quench upon which the temperature $T$ of the thermostat  would be lowered is meaningless: the quantity $\beta V_{SS}(r)$ is temperature-independent.
Instead, we quench by taking advantage of the fact that the functionality $f$ controls the softness of the potential $\beta V_{SS}(r)$ and thus for small $f$-values the stars are too penetrable to form a crystal or a glass: in this sense, the combination $f^{-3/2}$ plays a role similar to $T$ for thermal interactions, and therefore by choosing a sufficiently low $f$-value to quench from guarantees that the starting state will be a uniform, ergodic fluid.

We thus first fixed $f = 10$ and initialized all soft particles in a simple cubic lattice at $\rho_S\sigma_S^3 = 0.4$, equilibrating the system for a total of $10^3$ integration steps of size ${\rm d}t = 0.01\tau$ each.
At $f=10$ the system is guaranteed to be in a liquid state for the given star density,\cite{watzlawek:prl:1999} and thus we ensure there is no memory of the initial crystalline state.
Indeed, a liquid configuration ensues after the equilibration.
From this liquid configuration we begin quenching the system by gradually increasing $f$. 
We first raise the value of $f$ from $f=10$ in increments of 20 and perform $10^5$ integration steps for every new value of $f$, until we reach $f=90$. 
Throughout this quenching stage, the system always equilibrates into an ergodic fluid.
At $f=90$, dynamical slowdown increases and we run a total of $10^6$ integration steps for each $f$, while keeping increasing the functionality in steps of 10.
Finally, from $f=130$ to the target value of $f=166$ we run $10^7$ integration steps for each $f$-value. 
At the target value of $f = 166$, we then reach an equilibrated glassy state, as witnessed by two-step relaxation functions and mean-square displacements featuring the characteristic glassy plateau at intermediate times.

Next, we prepare seven different such glassy systems in which we add to each a different number of hard colloids to the equilibrated glassy liquid, namely $N_H = [200,300,400,500,600,700, 800]$, corresponding to the partial densities $\rho_H \sigma_S^{3} \cong [0.02,0.03,0.04,0.05,0.06,0.07,0.08]$. 
We initially placed the hard particles randomly into the simulation box. 
To avoid numerical instability caused by overlaps between particles, we first fix the interaction between the added hard colloids and all other particles to the bounded potential $\beta V_{\mathrm{equil}} (r) = A(1+\cos(\frac{\pi r}{c})), $with $A = 100$, $c=1.5 \sigma_S$ and $r$ the inter-particle distance. 
After the hard colloids are introduced we use a simulation step of ${\rm d}t = 0.005 \ \tau$ and run a total of $10^4$ integration steps with this bounded potential to remove overlaps. 
Finally, we reset the potentials to their proper expressions and run $5 \cdot 10^7$ integration steps to equilibrate the entire mixture. 
For each hard sphere density we prepare 10 independent realizations of the system by repeating this process (equilibration of glassy liquid and addition of hard colloids) ten times.
This allows us first of all to increase the statistics, and second to make an average over the disorder of the soft glassy state. 
Production runs start from these independent equilibrated configurations and run for a total of $10^7$ integration steps at all hard colloid partial densities $\rho_H$. 

\subsection{Evaluating structure and dynamics of the mixture}
\label{sec:methods:analysis}

We have calculated a number of structural and dynamic quantities characteristic of ergodic liquids and glasses. 
Equal-time correlators deliver information about the system structure: 
here we focus on the structure factors
\begin{equation}
    S_{ij}(q)=\delta_{ij}+\sqrt{\rho_i \rho_j} \, \hat{h}_{ij}(q), \quad \mathrm{with} \quad i,j \in \{H,S\}
    \label{eq:str}
\end{equation}
where $\hat{h}_{ij}(q)$ is the Fourier transform of $h_{ij}(r)=g_{ij}(r)-1$, and $g_{ij}(r)$ is the radial distribution function.~\cite{hansen2013theory} 
We anticipate that the mixture will, upon addition of colloids, approach a demixing transition, carrying along a strong enhancement of concentration fluctuations.
Thus, we also consider the composition-composition structure factor given by:\cite{bhatia1970structural,biben1991phase}
\begin{equation}
    S_{cc}(q)=x_H x_S^2 S_{HH}(q) - 2 (x_H x_S)^{3/2} S_{HS}(q)+x_H^2 x_S S_{SS}(q),
    \label{eq:str-cc}
\end{equation}
where $x_i \equiv N_i/(N_H + N_S)$, $i = H, S$, is the number fraction of component $i$ in the mixture.

The study of the dynamics of our mixture is based on different observables, which probe the system at different times. 
We characterize the distribution of displacements by means of the self-part of the van Hove function,~\cite{hansen2013theory}
\begin{eqnarray}
    G_s(\bm{r}, t) = \frac{1}{N}\left\langle \sum_{i=1}^N\delta \left(\bm{r}-\bm{r}_i( t)+\bm{r}_i(0)\right) \right\rangle, 
    \label{eq:self-vH}
\end{eqnarray}
which accounts for the fraction of particles out of the total number $N$ that performed a displacement $\Delta \bm{r}_i (t)=\bm{r}_i( t)-\bm{r}_i(0)$ over time $t$. 
Its moments of order $n$ can be calculated as
\begin{equation}
    \langle \Delta \bm{r}^n(t)\rangle = \frac{1}{N}\left\langle \sum_{i=1}^N  \left[\bm{r}_i(t) - \bm{r}_i(0)\right]^n \right\rangle. 
    \label{eq:MSD}
\end{equation}
In particular, the second moment, which identifies the Mean Squared Displacement (MSD), $MSD (t) \equiv \langle \Delta \bm{r}^2(t)\rangle$, together with the fourth moment, are used to define the non-Gaussian (NG) parameter
\begin{equation}
    \alpha_2(t)=\frac{3\langle \Delta \bm{r}^4(t)\rangle}{5\langle \Delta \bm{r}^2(t)\rangle^2}-1.  
    \label{eq:nongauss}
\end{equation}
Note that $\alpha_2(t)=0$ implies a Gaussian distribution of displacements.

The self- and collective Intermediate Scattering Functions (ISF) are given by
\begin{eqnarray}
    F_s(\bm{q},t)&=&\frac{1}{N}\left\langle \sum_{i=1}^N \exp\left[ \mathrm{i} \bm{q} \cdot (\bm{r}_i(t)-\bm{r}_i(0))\right] \right \rangle,  \label{eq:selfISF} \\
    F_c(\bm{q},t)&=& \frac{1}{N}\left\langle \sum_{i = 1}^N\sum_{j \ne i}^{N} \exp\left[ \mathrm{i} \bm{q} \cdot (\bm{r}_i(t)-\bm{r}_j(0))\right] \right \rangle, \label{eq:collectiveISF}
\end{eqnarray}
and mathematically represent the Fourier transform of the self- and collective van Hove functions.~\cite{hansen2013theory}
ISFs evaluate the relaxation times of the system and are quantities that can be related directly to scattering experiments. 
Note that, since our system is isotropic, all the quantities defined in this section are studied as functions of the radial coordinate only, $r=|\bm{r}|$, and the corresponding wave vector, $q=|\bm{q}|$.

\section{Results and Discussion}
\label{sec:res}
\subsection{Glass melting and incipient phase separation}
\label{sec:res:glass-melt-&-phase-sep}
\begin{figure}
    \centering
    \xincludegraphics[width=0.49\linewidth,label=a)]{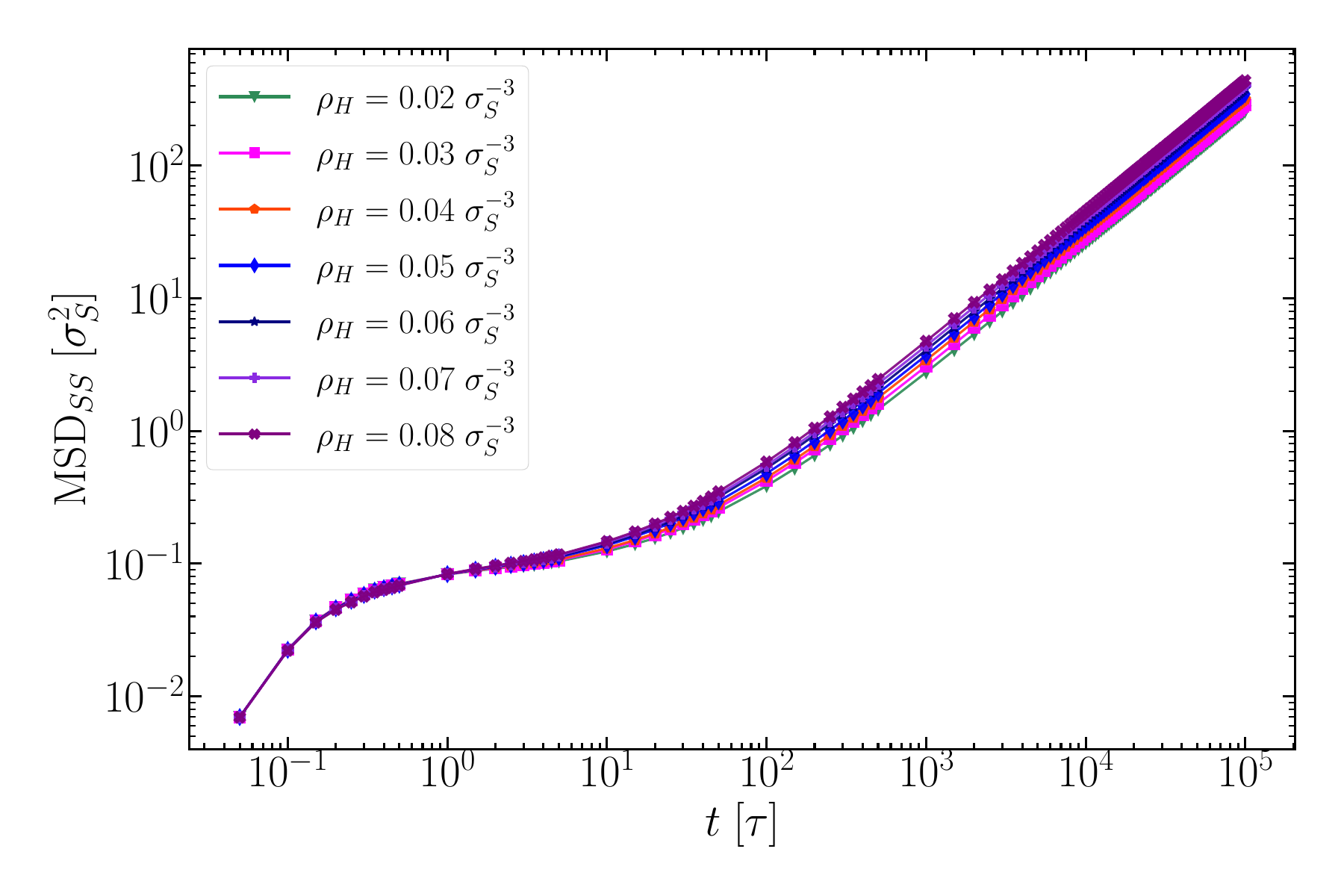}
    \xincludegraphics[width=0.49\linewidth,label=b)]{Fig3b.pdf}
    \xincludegraphics[width=0.49\linewidth,label=c)]{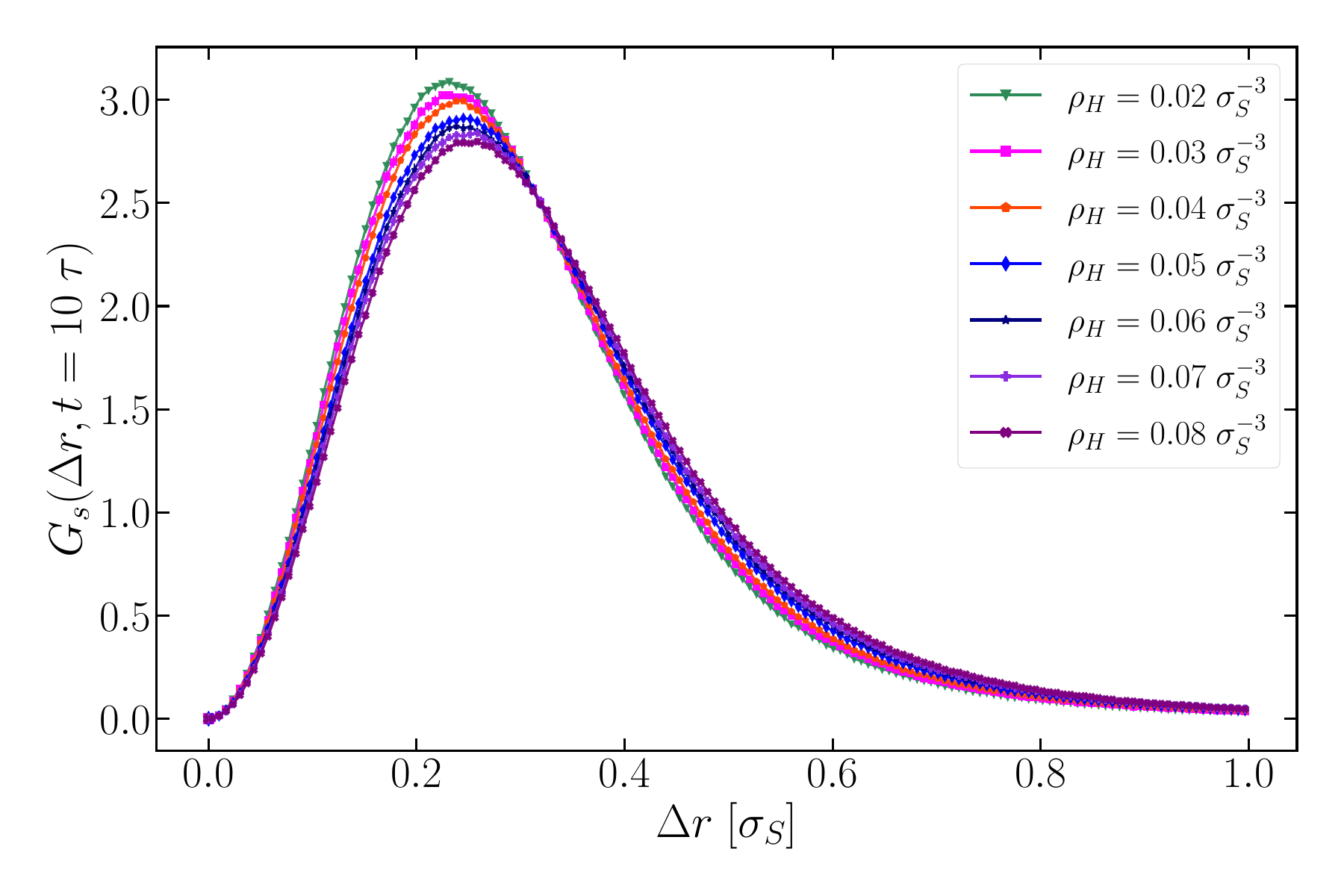}
    \xincludegraphics[width=0.49\linewidth,label=d)]{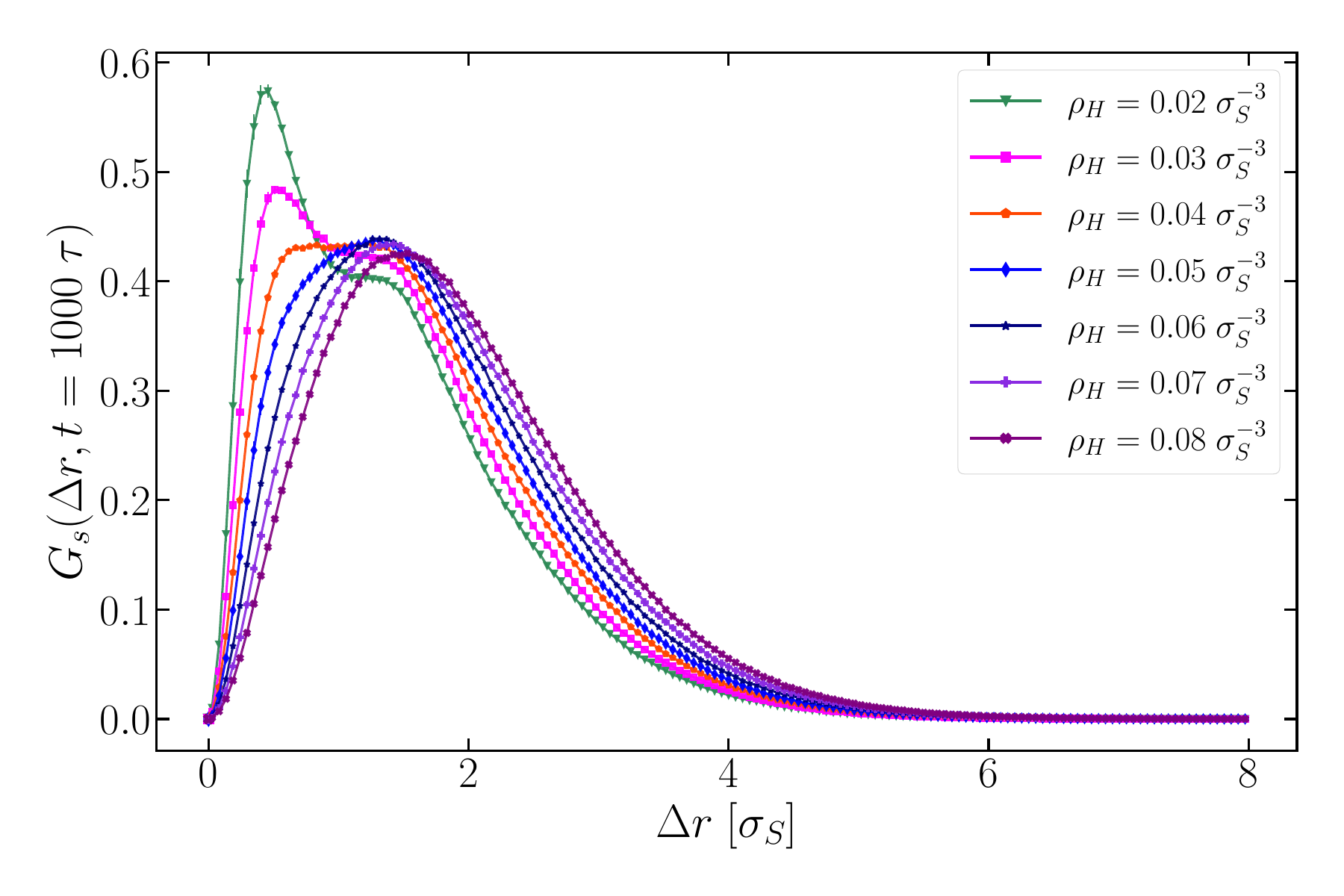}
    \caption{Dynamical quantities of the star polymer species at density
    $\rho_S\sigma_s = 0.4$, 
    upon addition of hard colloids with densities
    $\rho_H$ as indicate in the legends. (a) The mean square displacement 
    ${\mathrm {MSD}}_{SS}$ of the stars; (b) the self-ISF computed at $q = 4.99\,\sigma_S^{-1}$;
    (c) the self-van Hove functions computed at $t=10\tau$ and (d) 
    same as (c) at $t=1000\tau$. 
    The inset in panel (b) shows the values for the diffusion coefficient and the relaxation time as a function of the hard colloid density, rescaled by their reference value set at $\rho_H=0.02 \, \sigma_S^{-3}$.
    The former are obtained by fitting the MSD in panel (a) while the latter are obtained from the self-ISF curves as $F_s(q,\tau_\mathrm{rel}^S)=1/e$.}
    \label{fig:glass-melting}
\end{figure}

Various quantities characterizing the star polymer dynamics and related results are shown in Fig.~\ref{fig:glass-melting}. 
As discussed in section~\ref{sec:methods:simumlations}, we start from a pure star glassy system with a slow dynamics, characterized by a plateau both in the MSD and in the ISF, and add hard colloids. 
We note that upon adding colloids, the dynamics of the stars accelerates: the plateau shortens, both in the MSD [Fig.~\ref{fig:glass-melting}(a)] and the self ISF [Fig.~\ref{fig:glass-melting}(b)], the long-time diffusion coefficient increases, and the relaxation time decreases [Fig.~\ref{fig:glass-melting}(b), inset], in full agreement with experimental findings and MCT-results.~\cite{truzzolillo:prl:2013,marzi:sm:2015,parisi:jcp:2021}
Moreover, the self van Hove functions reveal a Gaussian distribution at short times, Fig.~\ref{fig:glass-melting}(c), while at longer times the resulting bimodal/non-Gaussian distribution seen in Fig.~\ref{fig:glass-melting}(d) indicates hopping and heterogeneous dynamics, as expected for a glassy system. 
Upon increasing $\rho_H$, the heterogeneous dynamics is suppressed, suggesting a restoration of ergodicity.
All these results indicate the melting of the soft glassy state, as expected with an increase in the hard sphere concentration, corresponding to the glass-liquid transition shown in the phase diagram sketched in Fig.~\ref{fig:phasediagr}.

\begin{figure}
    \centering
    \xincludegraphics[width=0.49\linewidth,label=a)]{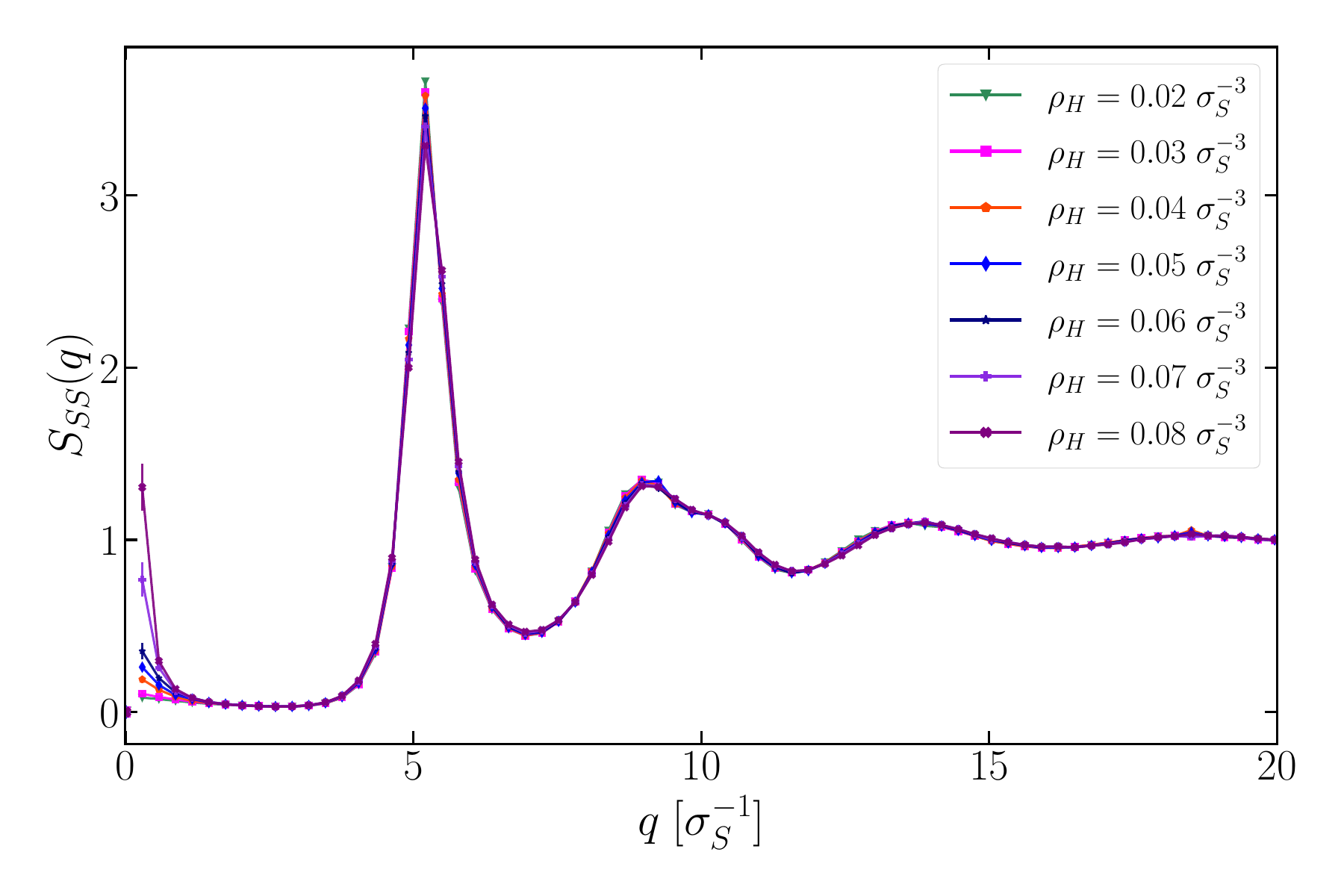}
    \xincludegraphics[width=0.49\linewidth,label=b)]{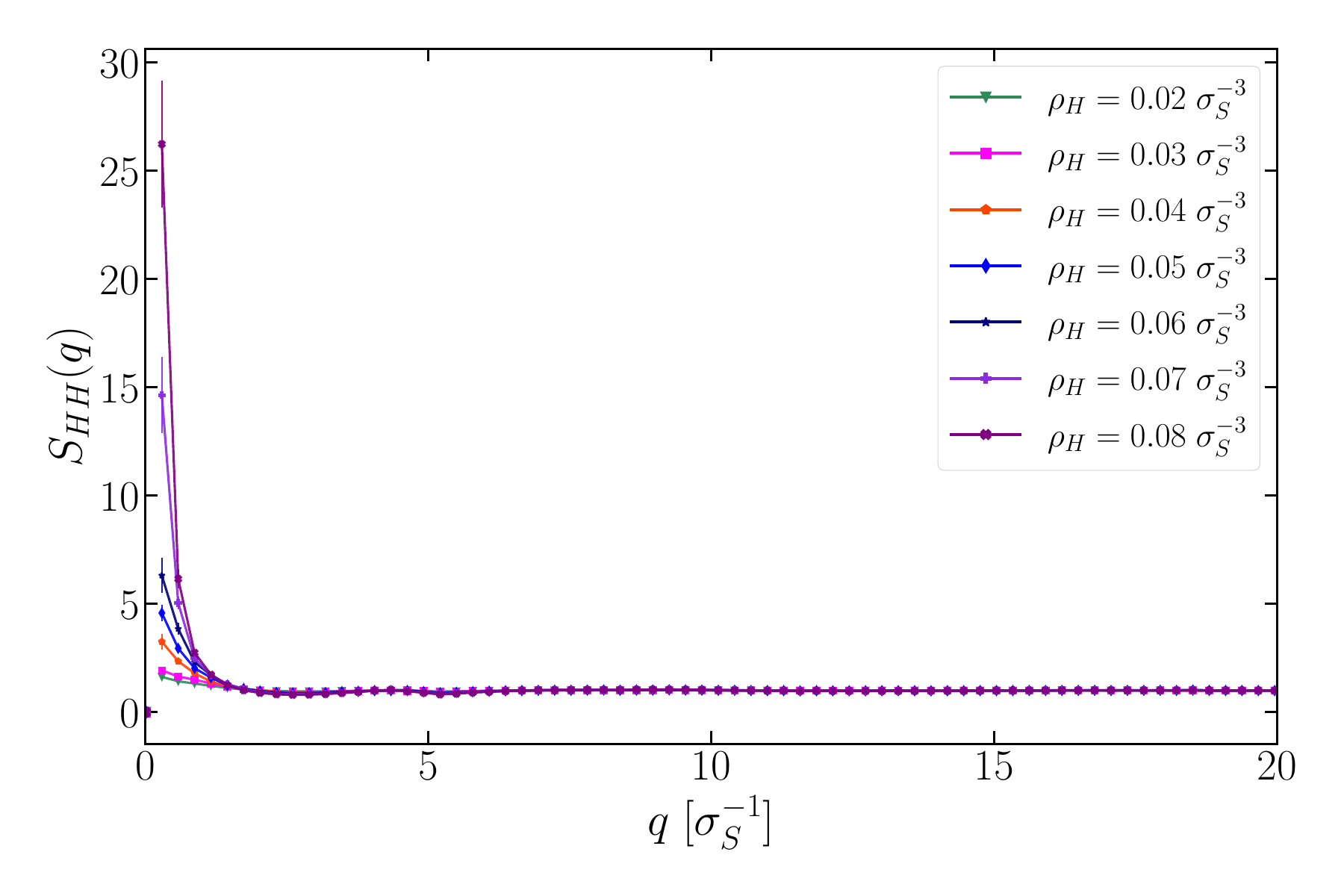} 
    \caption{(a) The star-star and (b) the hard sphere-hard sphere structure factors for different values of $\rho_H$, obtained from MD simulations.}
    \label{fig:strfact} 
\end{figure}

In Fig.~\ref{fig:strfact} we report the structure factor for both soft and hard components, $S_{SS}(q)$ and $S_{HH}(q)$ respectively, and its evolution with increasing hard colloid concentration. 
The effect of glass melting is much less pronounced if one looks at the structure factors; nevertheless, it can be seen in Fig.~\ref{fig:strfact}(a) that the height of the main peak of $S_{SS}(q)$ at $q \cong 5.2\,\sigma_S^{-1}$ decreases monotonically upon addition of colloids, a feature that is the main cause behind the MCT-transition from a glassy state with a nonvanishing non-ergodicity factor to an ergodic fluid upon addition of colloids.~\cite{marzi:sm:2015,parisi:jcp:2021}
Overall, the structure at intermediate- and large $q$-values is not 
markedly affected by changes in $\rho_H$. 
Moreover, as the colloid density is very low, the overall shape of the colloid structure factor $S_{HH}(q)$ in Fig.~\ref{fig:strfact}(b) is relatively featureless at finite $q$-values, especially in comparison with the star structure factor. 
However, both $S_{SS}(q)$ and $S_{HH}(q)$ start developing a peak around $q\to 0$ as $\rho_H$ grows.
This is indicative of increasing structural correlations at large scales, representative of an incipient phase separation, i.e., demixing, in full agreement with experimental results and integral equation theories.~\cite{truzzolillo:prl:2013,marzi:sm:2015,parisi:jcp:2021}
In fact, our results stress therefore that glass melting and incipient phase separation can coexist already at low hard sphere concentration, leading to interesting multi-scale phenomena as discussed below. 

\subsection{Hard colloid dynamics}
\label{sec:res:hard-sphere}

\begin{figure}
    \centering
    \xincludegraphics[width=0.49\linewidth,label=a)]{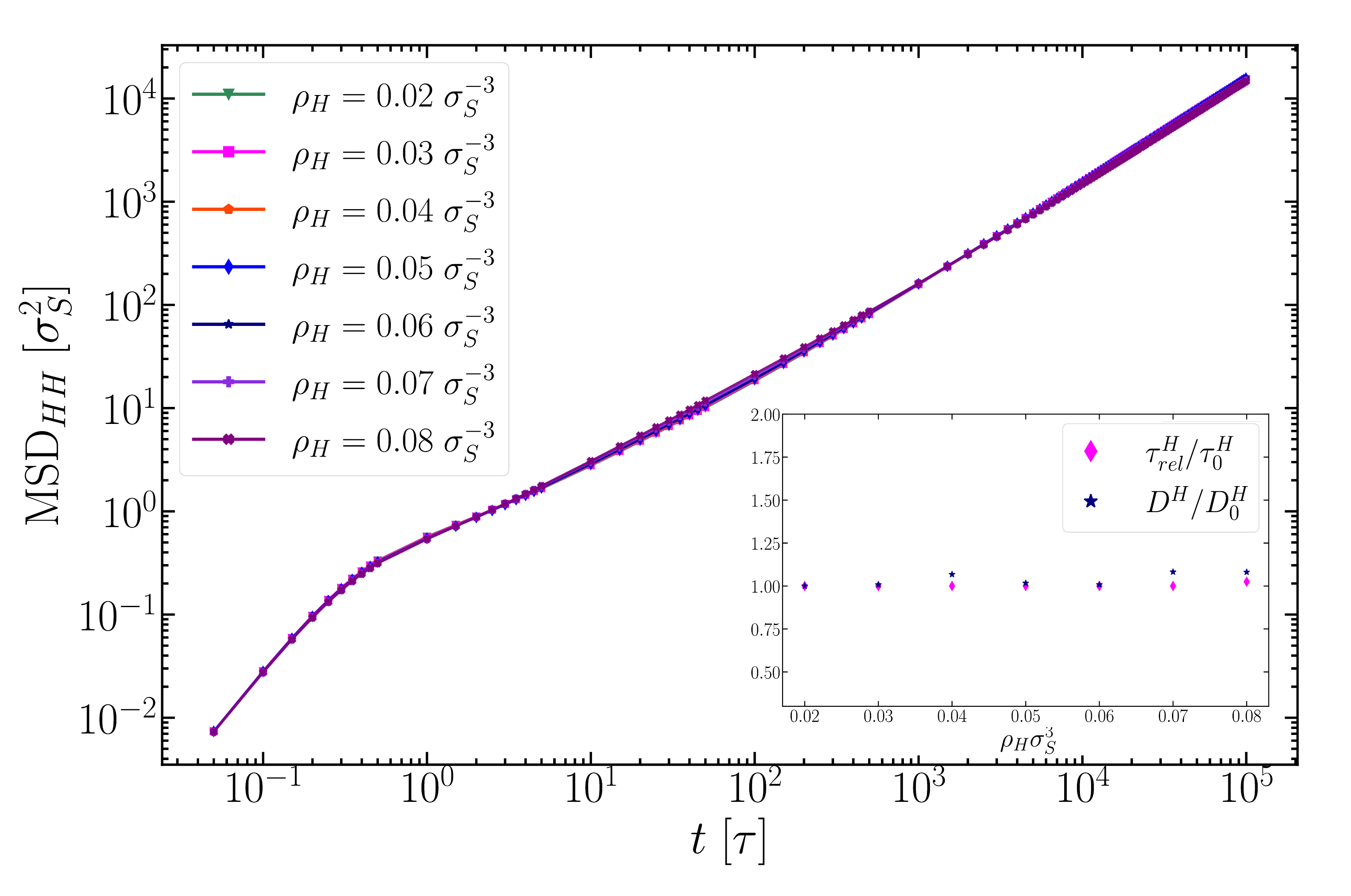}\\
    \xincludegraphics[width=0.49\linewidth,label=b)]{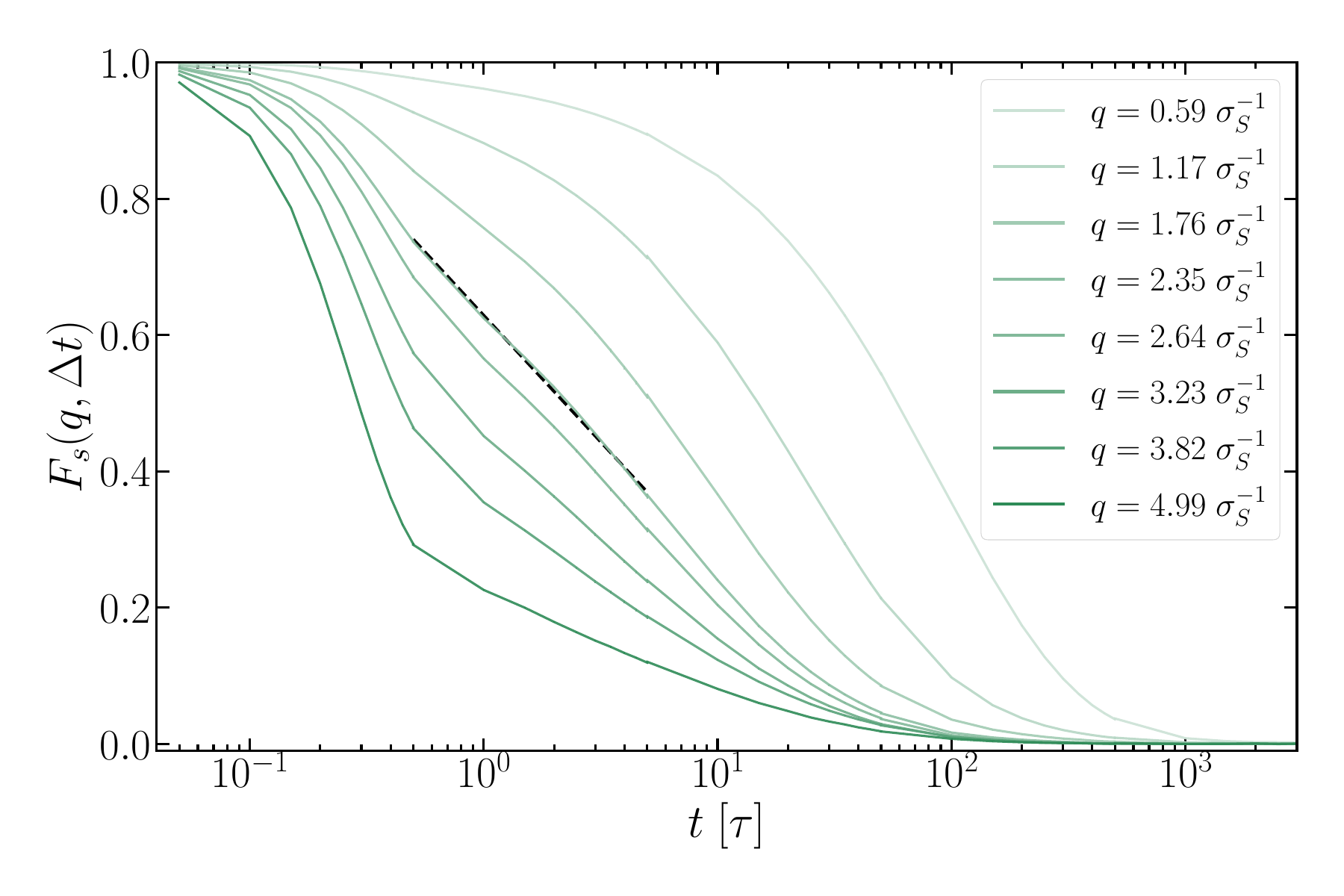}
    \xincludegraphics[width=0.49\linewidth,label=c)]{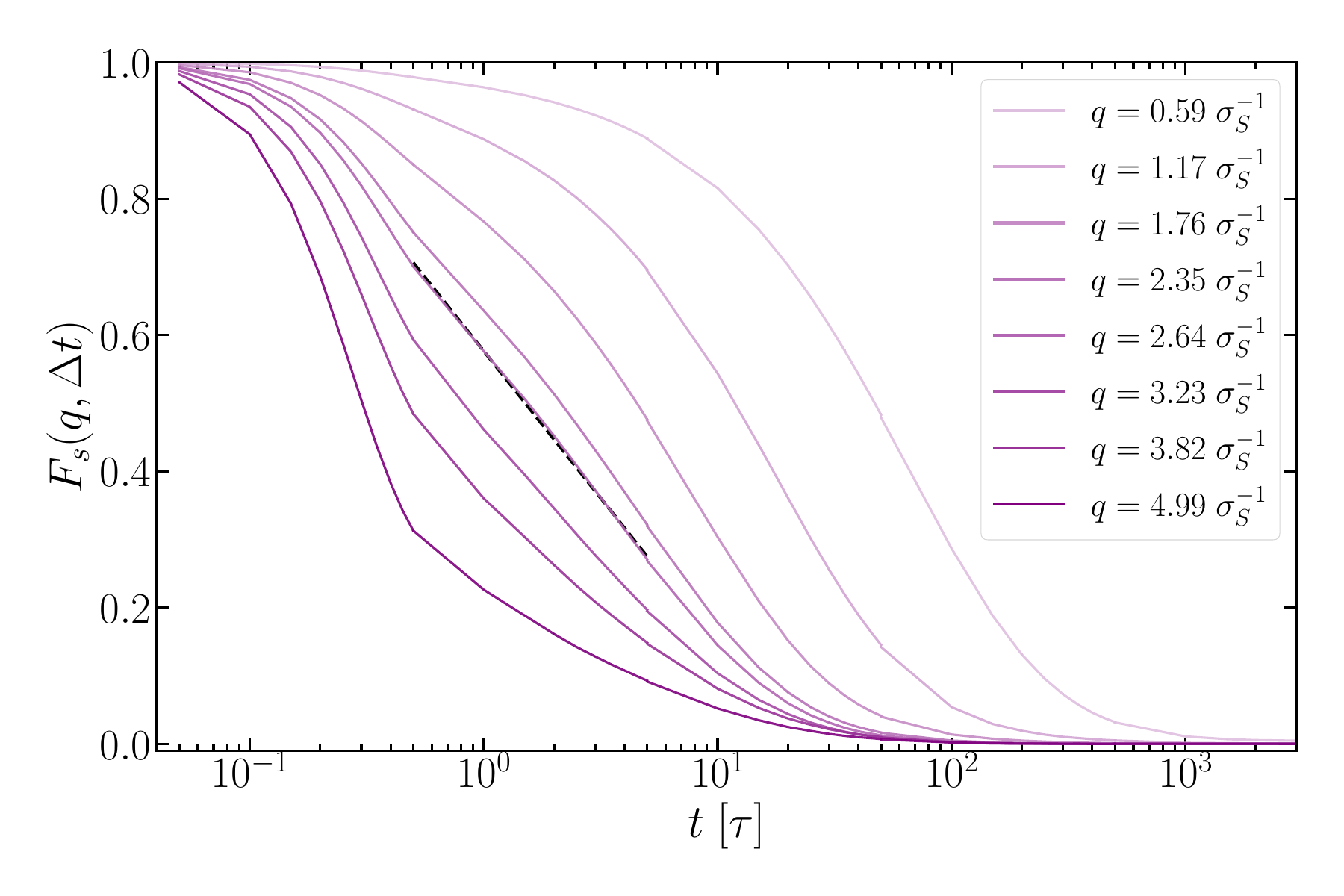}
    \caption{Hard colloid dynamics obtained from MD simulations. 
    (a) The ${\mathrm MSD}_{HH}$ of the hard component for different values of 
    $\rho_H$; the inset shows the values for the diffusion coefficient and the relaxation time as a function of the hard colloid density obtained as in Fig.~\ref{fig:glass-melting} and rescaled by their reference value set at $\rho_H=0.02 \, \sigma_S^{-3}$. 
    The self ISF for increasing $q$-values at two different concentration: (b) at $\rho_H=0.02 \, \sigma_S^{-3}$ and (c) at $\rho_H=0.08 \, \sigma_S^{-3}$. 
    The dashed black lines indicate a logarithmic fit, valid at $q_{\times}=2.35 \, \sigma_S^{-1}$ for $\rho_H=0.02 \, \sigma_S^{-3}$ and $q_{\times}=2.64 \, \sigma_S^{-1}$ for $\rho_H=0.08 \, \sigma_S^{-3}$.}
    \label{fig:HS-dynamics}
\end{figure}

Results on the dynamics of the added colloidal particles are shown in Fig.~\ref{fig:HS-dynamics}. 
The hard colloids show a diffusive MSD (following the initial ballistic regime), with an effective diffusion coefficient that is not affected by their concentration, as can be seen in Fig.~\ref{fig:HS-dynamics}(a) and its inset. 
On the basis of the MSD alone, no effect of the added colloids (on their own diffusion) can be seen, which contrasts the visible effect they have on the soft, glassy matrix, presented in the preceding subsection. 
However, there is a host of nontrivial effects present at the colloid tracer dynamics that cannot be captured by or reflected on their mean-square  displacement curves.

The self-ISF's of the hard component show a trend that is typical of tracers diffusing in a glassy matrix.~\cite{roberts2018tracer}
We focus in particular on two hard sphere densities, $\rho_H=0.02 \, \sigma_S^{-3}$ and $\rho_H  =0.08 \, \sigma_S^{-3}$, reported in Fig.~\ref{fig:HS-dynamics}(b) and Fig.~\ref{fig:HS-dynamics}(c), respectively, which show similar behavior. 
For very small wave vectors (or large length scales), the self-ISF decays exponentially, indicating diffusive behavior. 
At intermediate values of $q$ we start observing a logarithmic decay of the 
self-ISF at a crossover wavenumber $q_{\times}$, representing a competition between different relaxation dynamics.~\cite{dawson:pre:2000,mayer:mm:2009, chaudhuri2010gel, chaudhuri2015relaxation, sentjabrskaja2016anomalous,roberts2018tracer,Voigtmann:epl:2011,sciortino:prl:2003,gnan:prl:2014} 
In particular, we note that the logarithmic trend appears at higher $q_{\times}$ values for increasing $\rho_H$, suggesting that the typical length scale $\lambda_{\times} \propto 1/q_{\times}$ shortens as the soft matrix speeds up ($q_{\times}=2.35 \, \sigma_S^{-1}$ for $\rho_H=0.02 \, \sigma_S^{-3}$ and $q_{\times}=2.64 \, \sigma_S^{-1}$ for $\rho_H=0.08 \, \sigma_S^{-3}$). 
Finally, at large $q$-values (or small length scales), the self-ISF displays a sharp transition between two different exponential trends, the first one, at short times, representing the particle dynamics in the cage formed by the slow star matrix and the second one, at long times, corresponding to the exploration of the heterogeneous glassy matrix beyond the cage. 
The presence of the logarithmic decay is consistent with the existence of two nearby states of dynamical arrest: at the lower $\rho_H$-side, the star-polymer glass, which traps the colloidal tracers in its cages and thus introduces the aforementioned two-step decay of the self-ISF at short length scales.
The other glassy state is the arrested phase separation, which exists at higher $\rho_H$-values and it is caused by the simultaneous propensity of the added colloids to bring about a demixing transition and the vitrification line of the soft stars, which hinders this transition from fully materializing, creating large regions of inhomogeneous compositions that cannot relax to their
phase-separated equilibrium state.

\begin{figure}
    \centering
    \xincludegraphics[width=0.49\linewidth,label=a)]{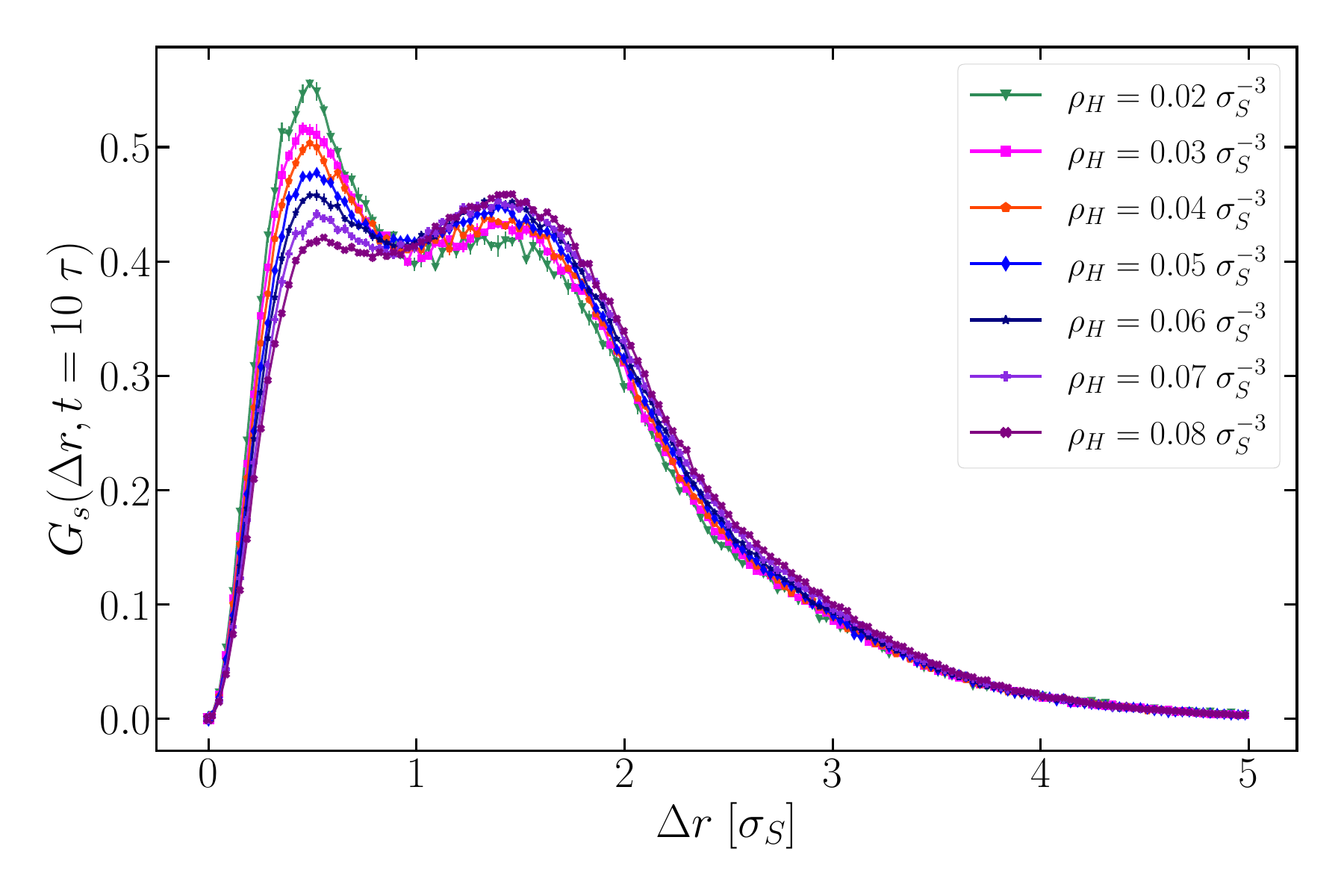}
    \xincludegraphics[width=0.49\linewidth,label=b)]{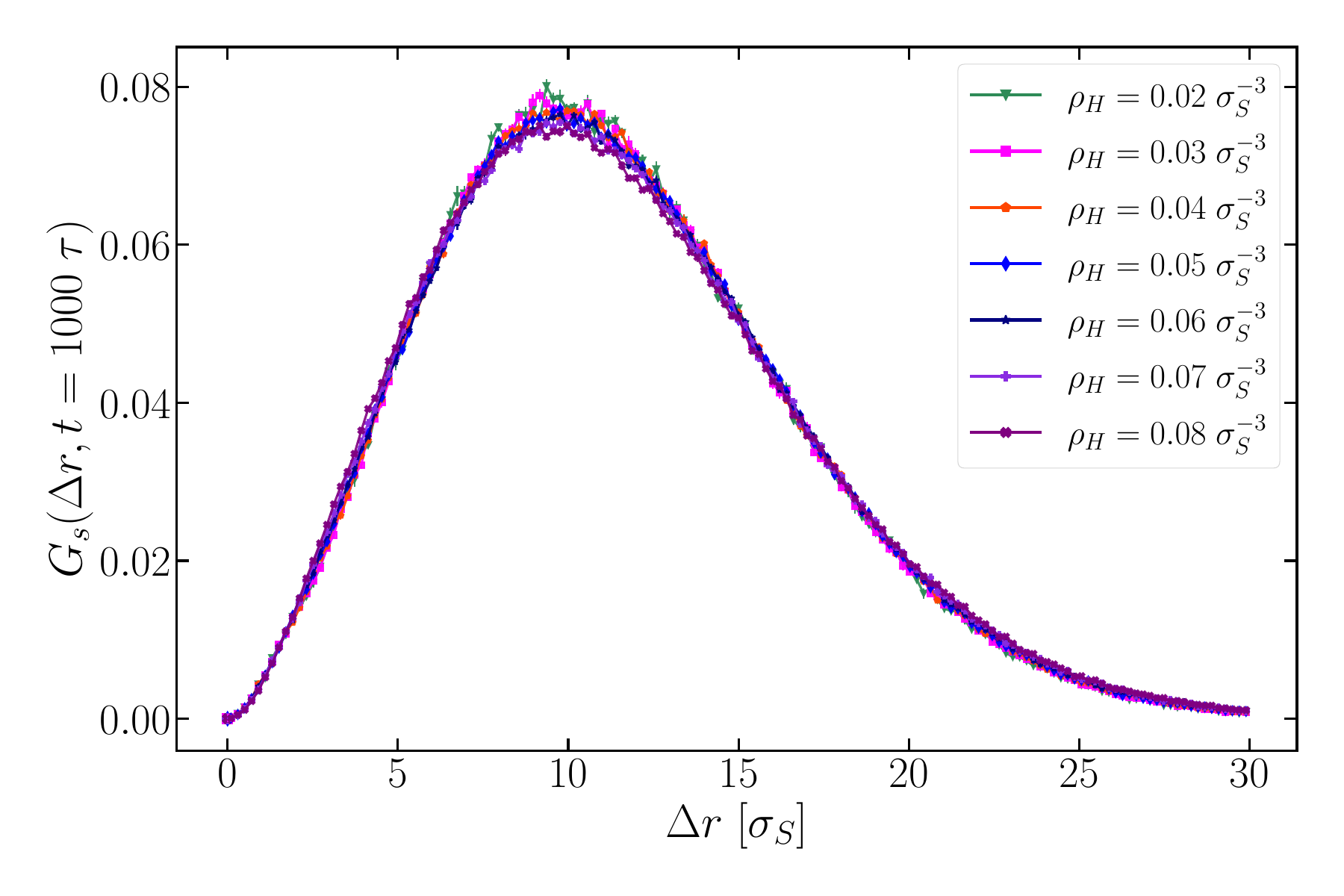}
    \xincludegraphics[width=0.49\linewidth,label=c)]{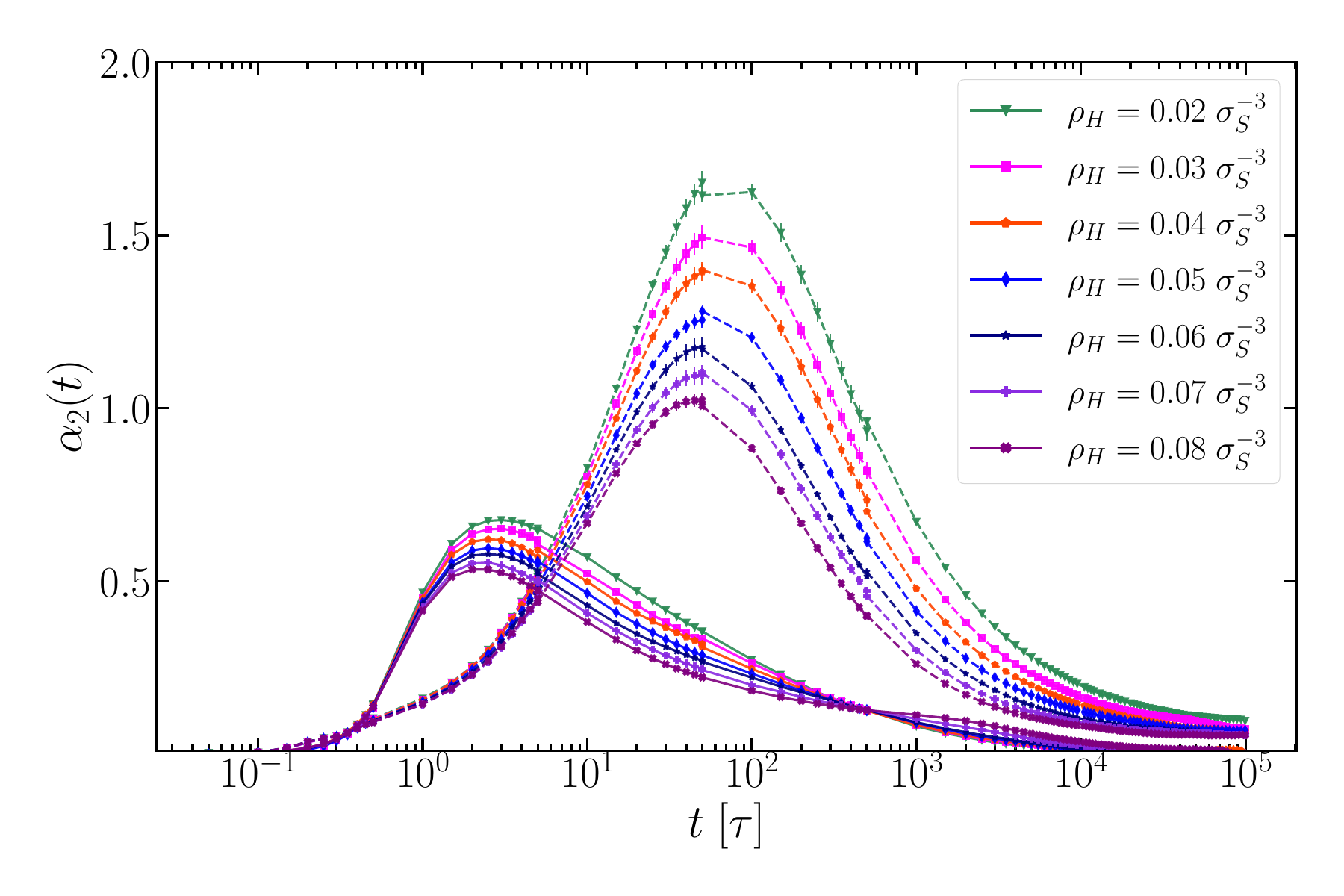}
    \caption{Deviations from Gaussianity of the hard components obtained from MD simulations. (a)-(b) Self van Hove functions of the hard components for different values of $\rho_H$ computed at two different time lags, (a) $\Delta t= 10 \tau$ and (b) $\Delta t= 1000 \tau$. (c) NG parameter of the hard component (solid lines) compared with the soft component one (dashed lines) for different values of $\rho_H$.}
    \label{fig:HS-vanhove-NG}
\end{figure}

Such trends of the self ISFs, if observed in real-space, correspond to deviation from Gaussianity of the displacement distribution, as seen in Figs.~\ref{fig:HS-vanhove-NG}(a)-and \ref{fig:HS-vanhove-NG}(b), which have features that are specular to those seen for the stars' counterpart shown in Figs.~\ref{fig:glass-melting}(c) and \ref{fig:glass-melting}(d). 
In particular, the displacement distribution seems to converge to a Gaussian shape common to all $\rho_H$ values at long times whereas, at short times, the hard colloids clearly display a bimodal displacement distribution function. 
We can quantify these deviations from Gaussianity by looking at the NG parameter $\alpha_2(t)$, Eq.~(\ref{eq:nongauss}), which is shown in Fig.~\ref{fig:HS-vanhove-NG}(c), both for the soft and hard components.
The NG parameter of the hard component is in line with what is observed in the literature for tracers diffusing in a glassy matrix,~\cite{roberts2018tracer}
at least at short and intermediate times. 
We notice in particular that with decreasing $\rho_H$ the first peak of $\alpha_2(t)$ increases both for the soft and the hard components, showing that a pronounced glassy dynamics of the matrix leads to a more heterogeneous dynamics for the hard colloids as well. 
Indeed, the soft component displays a NG parameter typical of a glassy system: it starts from zero at short times, develops a peak at intermediate times, and relaxes back to zero at long times. 
The peak is stronger for smaller concentrations of hard spheres; the more hard spheres, the less glassy the system is, as stressed in section~\ref{sec:res:glass-melt-&-phase-sep}.
However, the NG parameter of the hard component also shows a second peak at longer times, with a trend with $\rho_H$ that is inverted to what was just described; that is, the higher $\rho_H$ the higher the peak. 
This suggests that there is a second source of non-Gaussianity in the dynamics of the hard colloids that is not directly linked to the glassiness of the matrix. 
We argue that such source is the population splitting that emerges as a consequence of the arrested phase separation, which grows together with $\rho_H$.

\begin{figure}    
    \xincludegraphics[width=0.49\linewidth,label=a)]{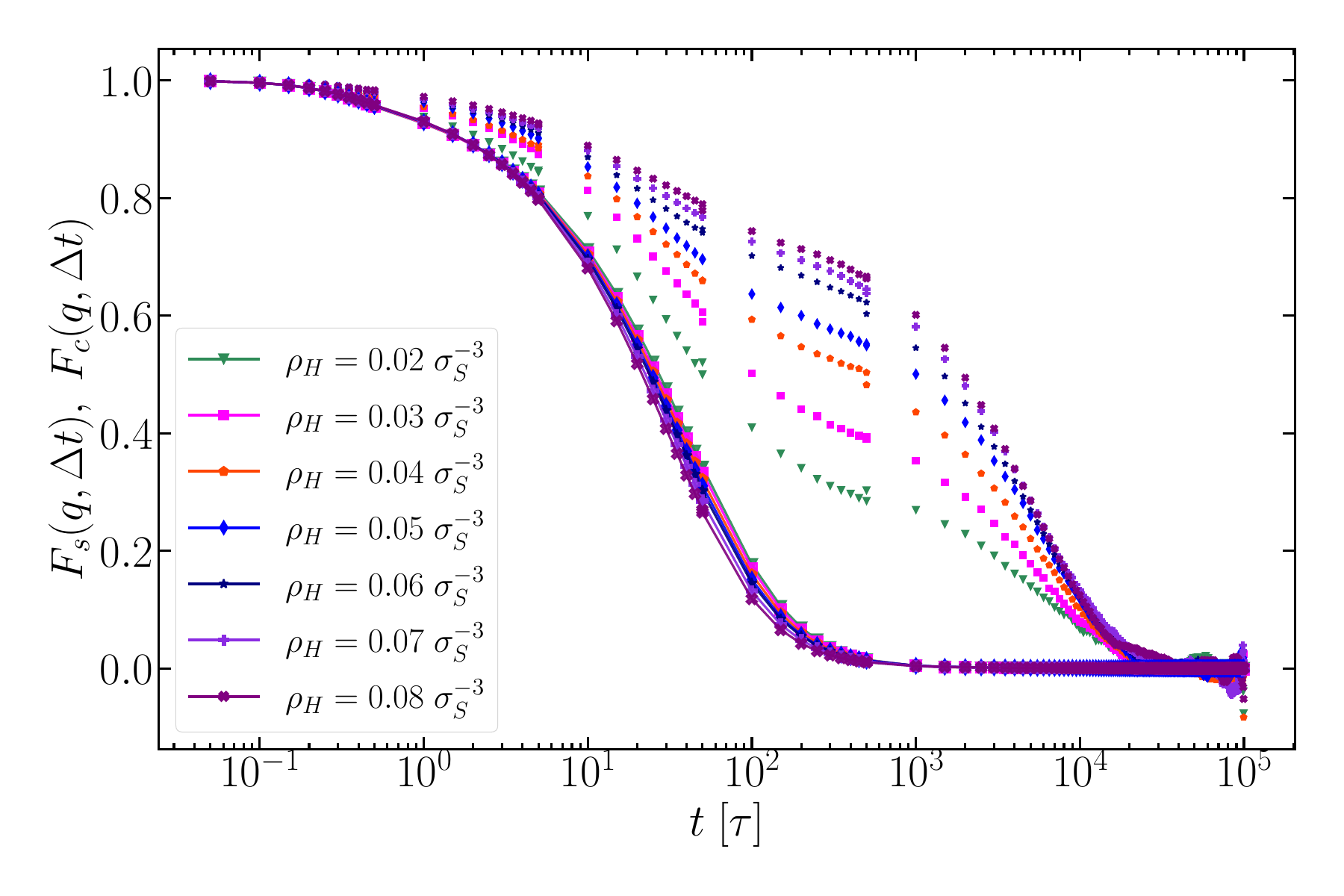}
    \xincludegraphics[width=0.49\linewidth,label=b)]{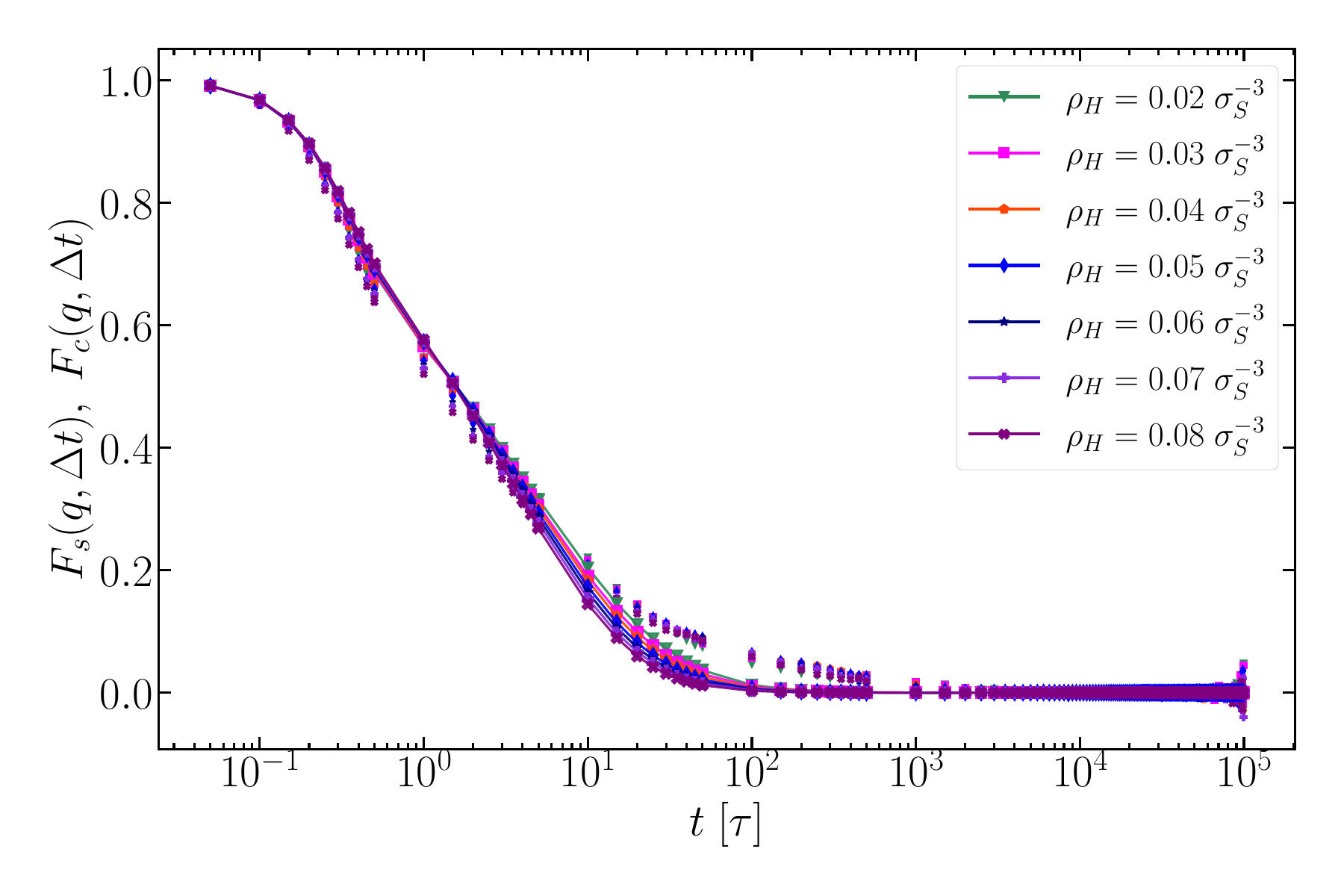}\\
    \xincludegraphics[width=0.49\linewidth,label=c)]{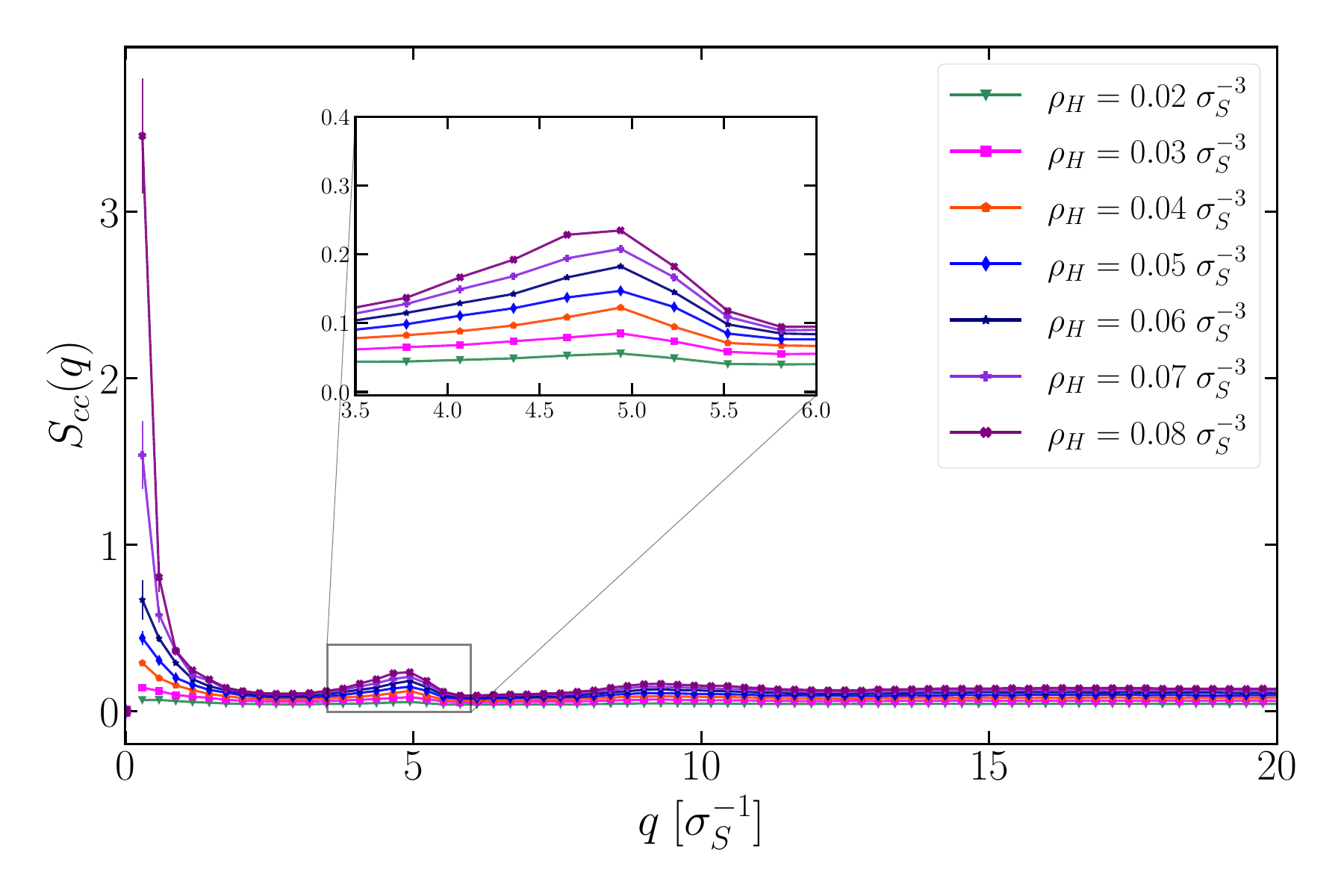}
    \caption{(a)-(b) Self (solid lines) and collective (markers only) intermediate scattering functions for hard components for different $\rho_H$ values computed at (a) $q=0.88\,\sigma_S^{-1}$ and (b) $q=2.64\,\sigma_S^{-1}$. 
    (c) Composition-composition structure factor for different values of $\rho_H$; the inset highlights the emergence of the second peak. }
    \label{fig:ISF-self-collective-Scc}
\end{figure}

The presence of arrested phase separation influences not only the single-particle dynamics but also the collective dynamics of the hard component. 
In Fig.~\ref{fig:ISF-self-collective-Scc} we report both the self-- and collective ISFs for the hard colloids at two different wave vector values, namely $q=0.88\,\sigma_S^{-1}$ and $q=2.64 \,\sigma_S^{-1}$.
As shown in Fig.~\ref{fig:ISF-self-collective-Scc}(a), for small values of $q$ there is a decoupling between the self- and collective dynamics of the hard component, which is increasingly evident as $\rho_H$ increases. 
We interpret this as evidence of growing phase separated regions of hard colloids, originating from the glassy environment in which the colloids move. 
Indeed, the correlated movements of groups of particles in regions of size $\sim q^{-1}$ are influenced by the slow dynamics of the majority component much stronger than individual particle movements and thus diffusion on the length scale $\sim q^{-1}$ becomes slow. 
The origin of this decoupling at the arrested phase separation is corroborated by three features:
first, the degree of decoupling grows with $\rho_H$, despite the fact that the star dynamics is accelerated by this addition. 
This clearly implies that there are large regions of strong composition inhomogeneities, so that the colloids that are trapped in the domains of high $\rho_S$ collectively relax slowly. 
Second, all collective ISFs relax to zero at the same time, despite the difference in the plateau heights, implying that the relaxation time is determined by the relaxation of the phase-separated denser star regions alone. 
And third, the decoupling appears for low $q$-values, signaling length scales vastly exceeding the size of individual particles, cages or small clusters thereof.

At larger $q$-values (small length scales), the decoupling disappears, as can be seen in Fig.~\ref{fig:ISF-self-collective-Scc}(b).
To assess the meaning of the structural length scales involved, we show in Fig.~\ref{fig:ISF-self-collective-Scc}(c) the composition-composition structure factor $S_{cc}(q)$, which encodes the significance of composition fluctuations at scales $\sim q^{-1}$. 
The dominant feature is indeed an incipient divergence at $q \to 0$ as $\rho_H$ increases; however, as seen in the inset of Fig.~\ref{fig:ISF-self-collective-Scc}(c), a maximum at finite $q \cong 5 \, \sigma_S^{-1}$ emerges as well, pointing to the appearance of composition fluctuations at scales $\lambda \cong (2\pi)/q \cong 1.3\,\sigma_S \cong 1.9 \sigma_H$.
This peak corresponds to the formation of local dense colloidal regions on top of the long-wavelength composition fluctuations, which can be seen in Figs.~\ref{fig:locdens}(c) and \ref{fig:locdens}(d). 
However, the decoupling between the self- and collective ISF of the colloids disappears already at 
smaller $q$-values, therefore its existence at low $q$ is a consequence of the arrested phase separation alone.

\subsection{Demixing and population splitting}
\label{sec:res:demixing}

To quantify the structure and dynamics of the mixture as it approaches phase separation in the vicinity of an arrest line, we introduce the instantaneous local density $\rho^i_\mathrm{loc}(t)$ around a given hard colloid $i$ as 
\begin{equation}
    \rho^i_\mathrm{loc}(t) = \frac{3}{4 \pi r_c^3} \sum_{j\ne i}^{N_H} \Theta (r_c- r_{ij}(t)),
    \label{eq:locdensity}
\end{equation}
where $\Theta(r)$ is the Heaviside step function, $r_{ij}(t)$ represents the distance of the particle $i$ from the particle $j$ at time $t$, and $r_c$ defines the neighborhood radius of a hard colloid. 
We tested different values of $r_c$ and found that $r_c=2.5 \sigma_S$ is an appropriate value to identify the phase separated regions.
The probability of observing a given local density $\rho_\mathrm{loc}$ over a whole trajectory in the entire population of hard colloids is given by
\begin{equation}
    P(\rho_\mathrm{loc})=\frac{1}{N_H} \sum_{i=1}^{N_H} 
    \frac{1}{{\mathcal N_f}}\sum_{\alpha=1}^{{\mathcal N}_f} \delta \left( \rho_\mathrm{loc} - \rho^i_\mathrm{loc}(t_{\alpha})\right),
    \label{eq:p-locdensity}
\end{equation}
where $N_H$ is the number of hard colloids in the system, ${\mathcal N_f}$ is the number of frames in the trajectory and $t_{\alpha}$ is the simulation time at a frame $\alpha$.
\begin{figure}
    \centering
    \xincludegraphics[width=0.49\linewidth,label=a)]{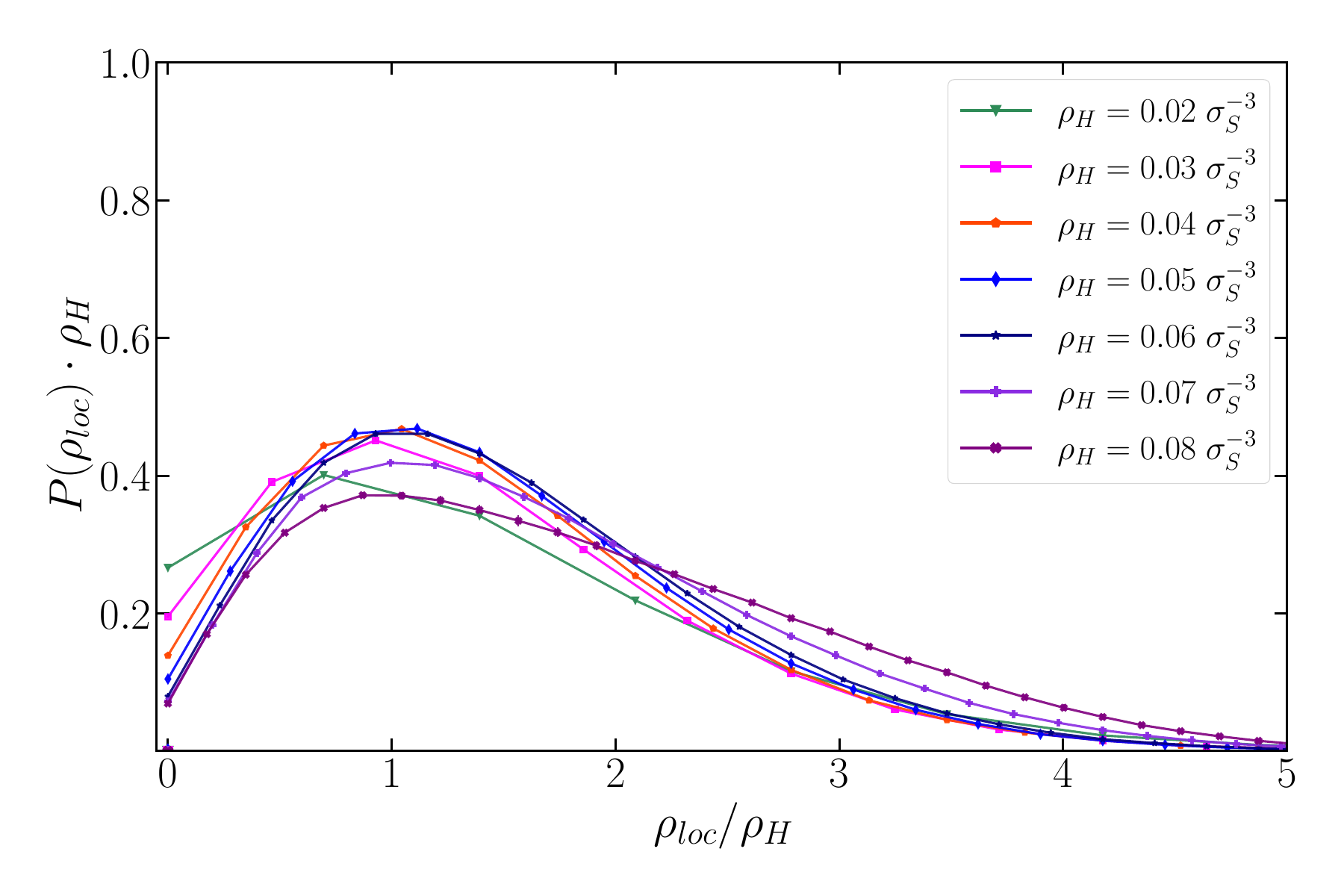}
    \xincludegraphics[width=0.49\linewidth,label=b)]{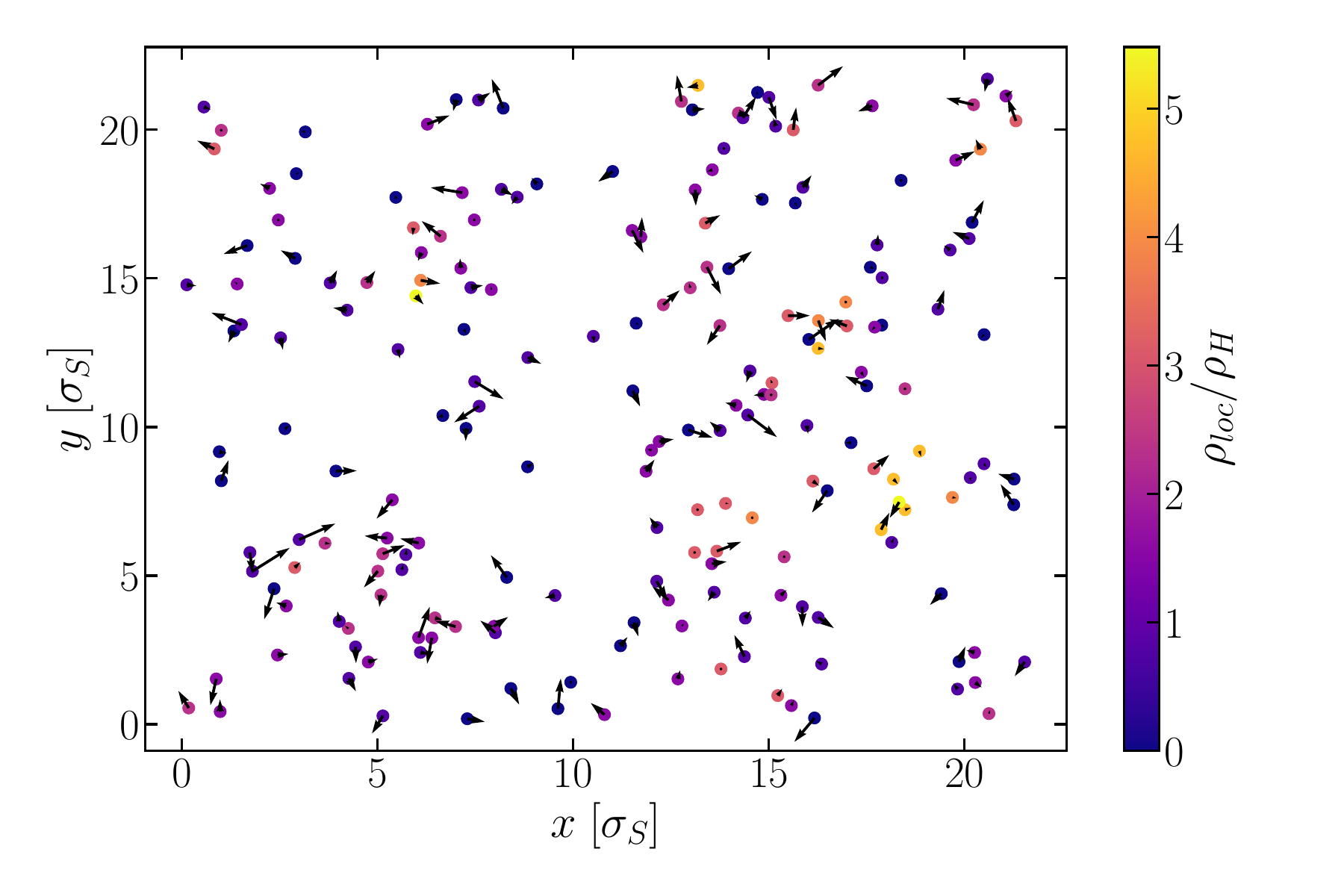}
    \xincludegraphics[width=0.49\linewidth,label=c)]{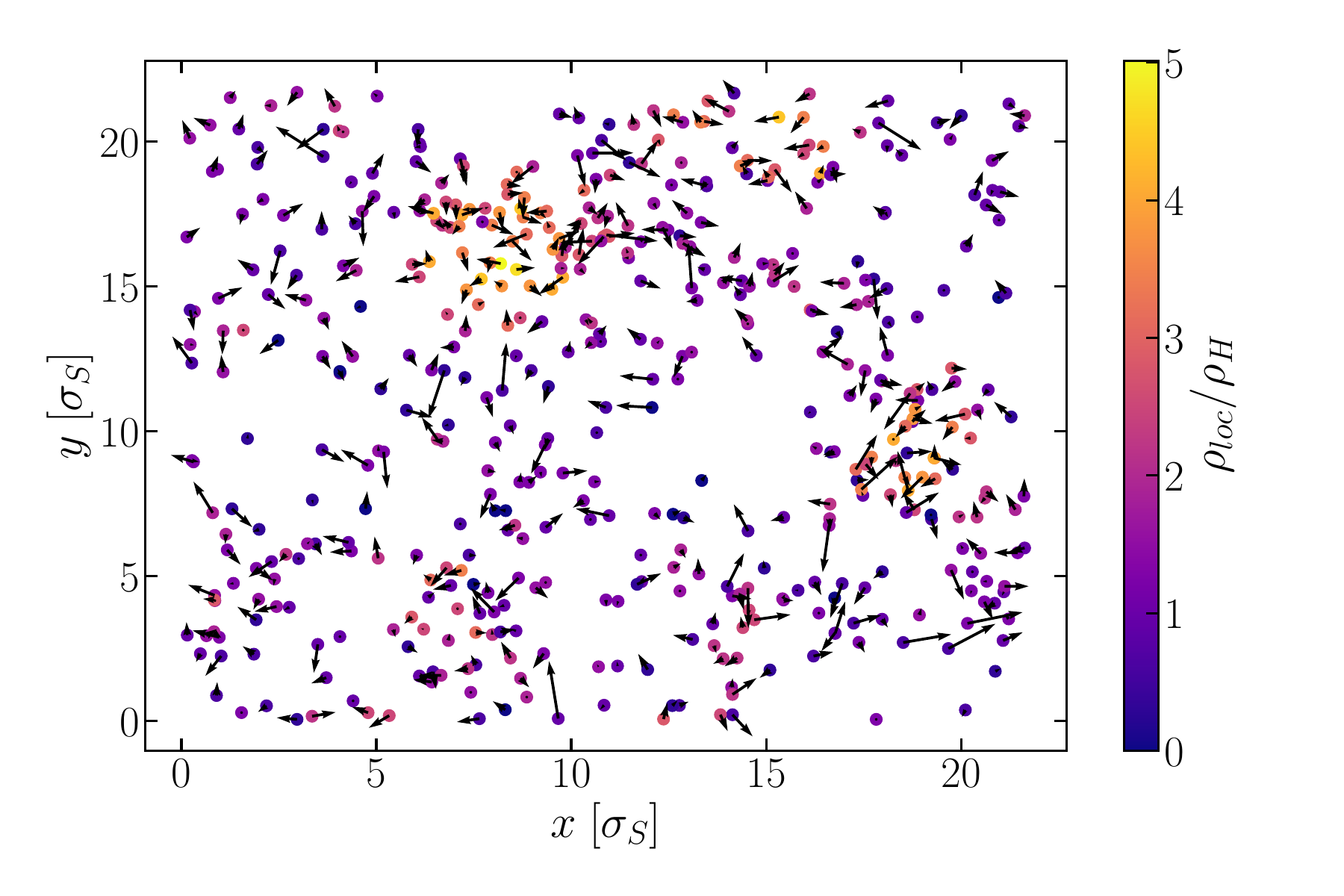}
    \xincludegraphics[width=0.49\linewidth,label=d)]{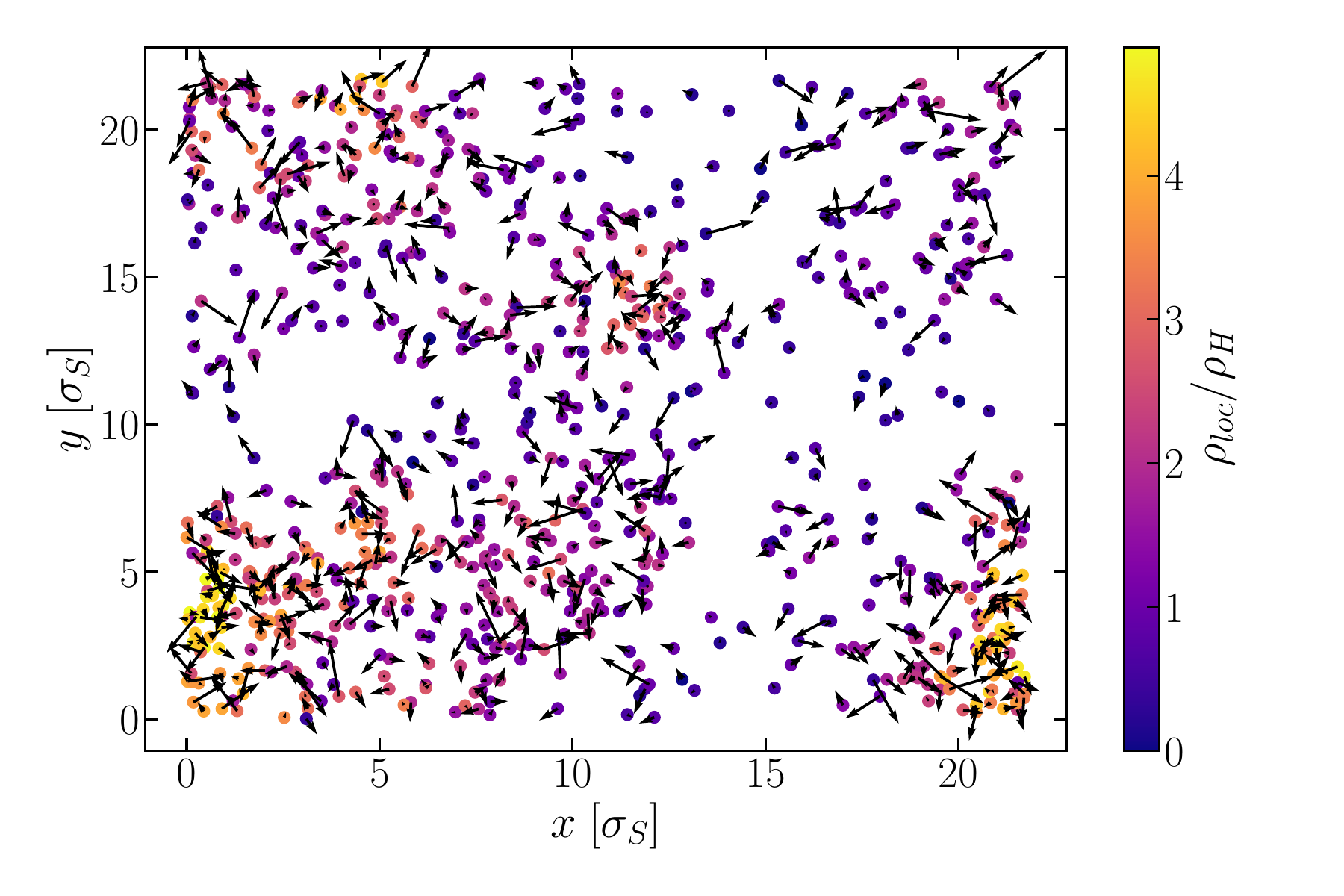}
    \caption{(a) Local density of hard colloids as defined in Eq.~\eqref{eq:locdensity} for different $\rho_H$ values. 
    In order to compare the different curves, the local density is normalized by the corresponding bulk density $ \rho_H$. 
    (b)-(d) Two-dimensional cuts of the system snapshots for (b) $\rho_H=0.02 \, \sigma_S^{-3}$, (c) $\rho_H=0.05 \, \sigma_S^{-3}$, and (d) $\rho_H=0.08 \, \sigma_S^{-3}$. 
    The arrows indicate the displacement calculated for each particle within a time lag $\Delta t = 10 \, \tau$. 
    Note that we are plotting the hard colloids only; the stars are ommitted
    for clarity.}
    \label{fig:locdens}
\end{figure}

\begin{figure}
    \centering
    \xincludegraphics[width=0.49\linewidth,label=a)]{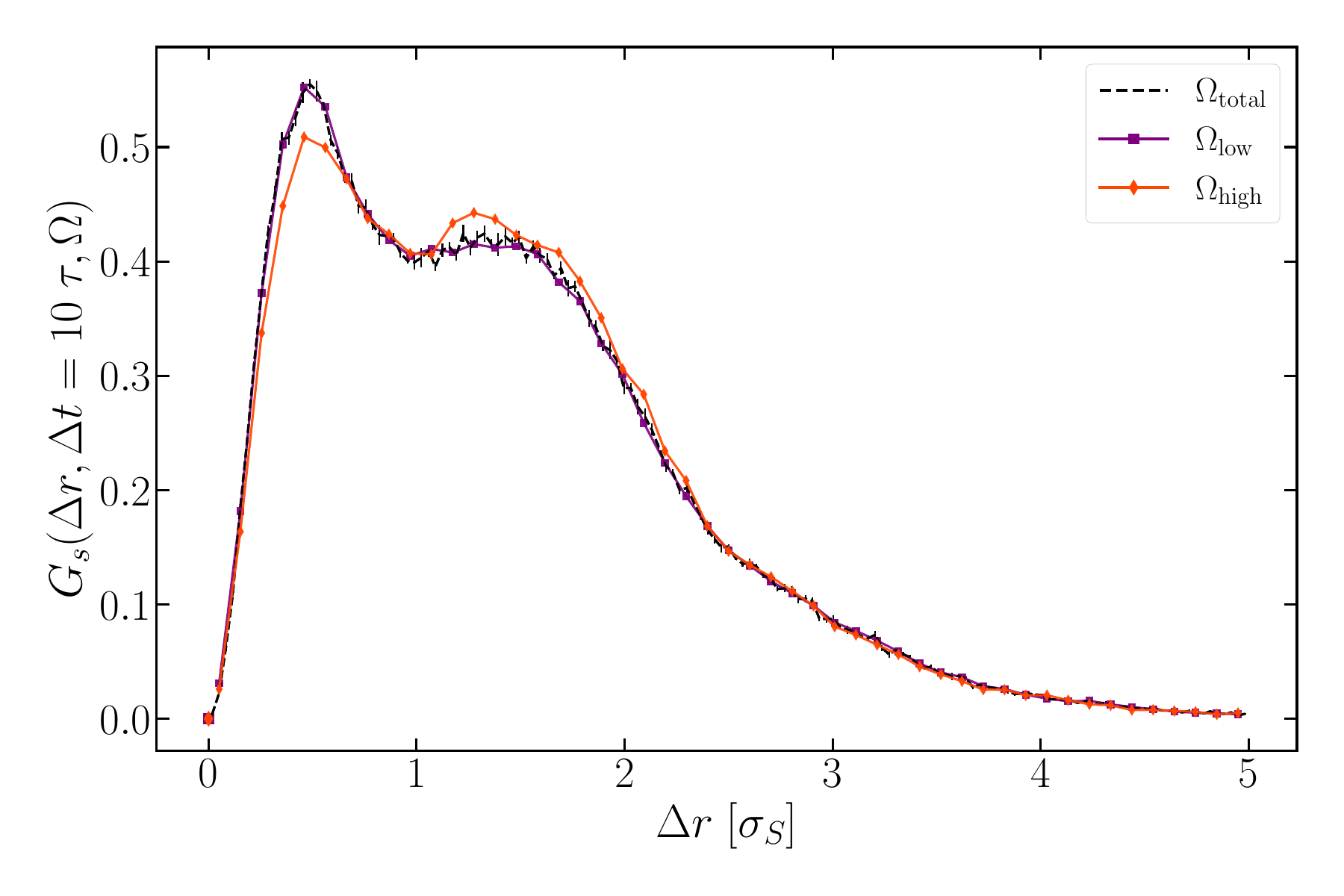}
    \xincludegraphics[width=0.49\linewidth,label=b)]{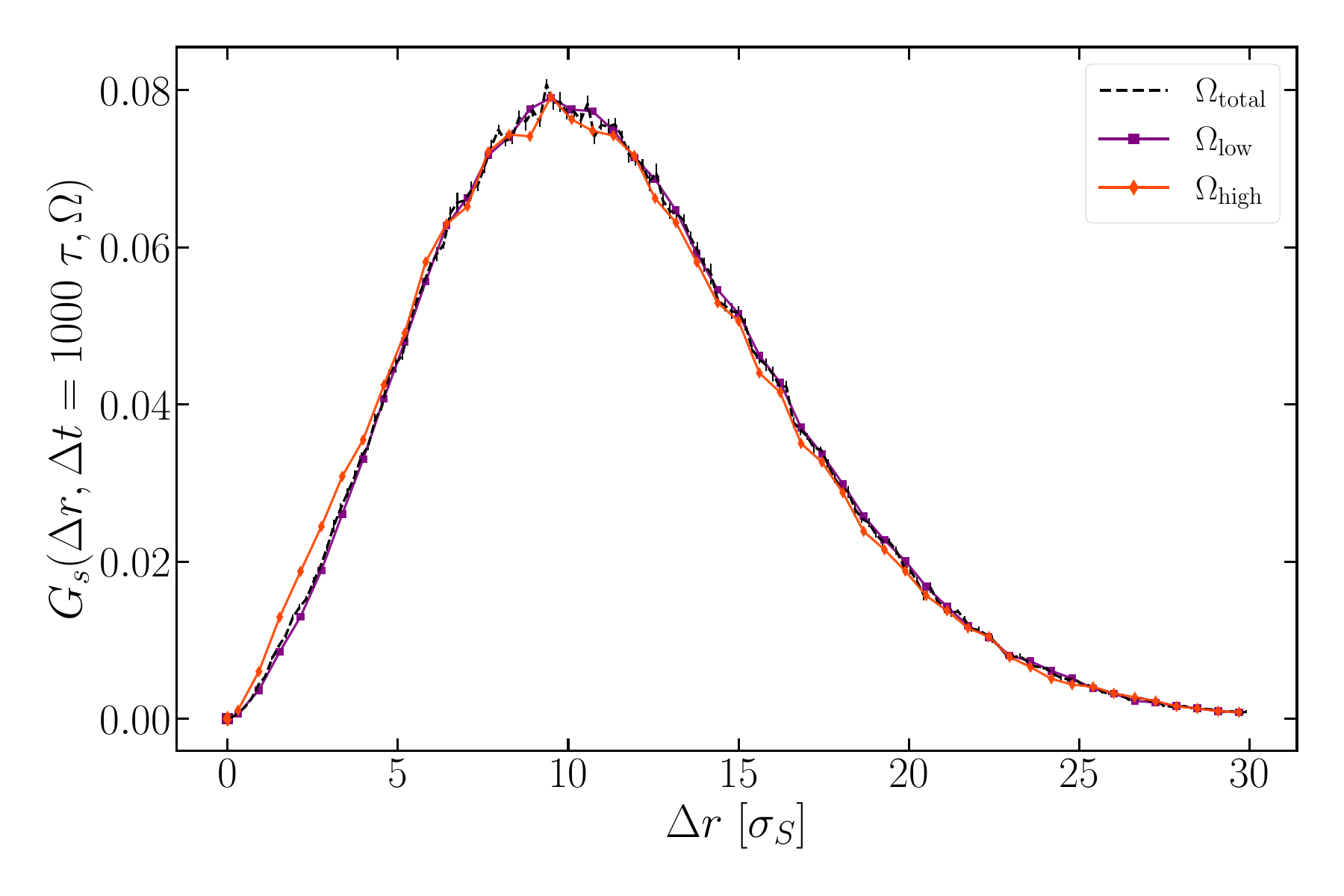}
    \xincludegraphics[width=0.49\linewidth,label=c)]{FIg9c.pdf}
    \xincludegraphics[width=0.49\linewidth,label=d)]{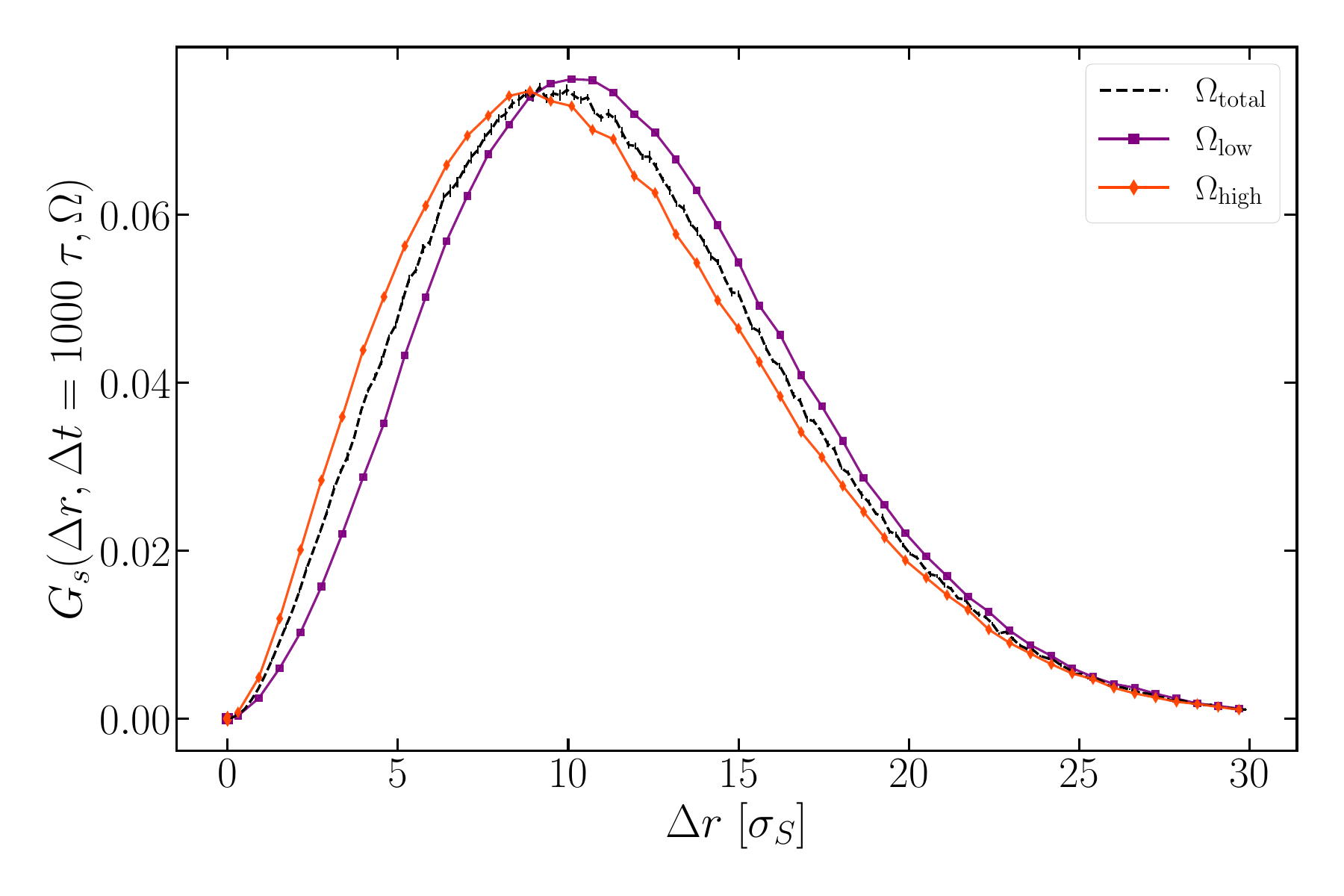}
    \caption{Self van Hove functions for the hard component split into contributions from different local density regions, evaluated at $\Delta t = 10 \,\tau$ and $\Delta t = 1000 \,\tau$ for two different hard sphere densities: (a)-(b) $\rho_H=0.02 \, \sigma_S^{-3}$,  and (c)-(d) $\rho_H=0.08 \, \sigma_S^{-3}$}
    \label{fig:vanhove-loc}
\end{figure}

In Fig.~\ref{fig:locdens}(a), we show the distribution of local density normalized over the bulk density. 
It can be seen that upon increasing $\rho_H$, the local density distribution displays a fatter tail, indicating the presence of more and more colloid rich regions. 
This feature can be clearly seen in the two-dimensional-cut of the system's snapshots reported in Figs.~\ref{fig:locdens}(b)-(d). 
Here, we plot the colloids, color-coded by their local density, and their corresponding displacements performed within a time interval $\Delta t = 10 \, \tau$, indicated by the arrows. 
We notice that the displacements are larger in the colloid rich regions while they are smaller in the colloid poor regions. 
This suggests that the arrested phase separation also leads to a population splitting of faster and slower colloids.
In line with this observation, we define the constrained displacement distribution for a given lag time $\Delta t$, conditional on the region $\Omega$ in which the colloids are found:
\begin{equation}
    G_s(r, \Delta t, \Omega)= \frac{1}{N}\left\langle \sum_{i=1}^N\delta \left(r-|{\bm r}_i(\Delta t)-{\bm r}_i(0)|\right)\, \chi\left( \rho_\mathrm{loc}/\rho_H , \Omega \right) \right\rangle,
    \label{eq:selfvH-cluster}
\end{equation}
where the function $\chi(x,\Omega)$ selects particles in a dimensionless scaled density interval as follows
\begin{equation}
    \chi(x,\Omega)=
    \begin{cases}
        1, & x\in \Omega, \\
        0, & \mathrm{otherwise}.
    \end{cases}
\end{equation}
Using $\Omega_\mathrm{low}= [0,1.5]$ for the colloid-poor region, and $\Omega_\mathrm{high}= [1.5,4.5]$ for the colloid-rich region, we decompose the total displacement distributions $G_s(r,t)$ into contributions from distinct environments, isolating the effect of phase separated regions ($\Omega_\mathrm{total} = [0,\infty)$ identifies the whole region). 
The results of this analysis are reported in Fig.~\ref{fig:vanhove-loc}.
From these plots we can indeed see that the small displacement peak comes mainly from colloids that started in a region $\Omega_\mathrm{low}$, while the larger displacement one is dominated by colloids that started in a region $\Omega_\mathrm{high}$; 
the effect is more pronounced for increasing $\rho_H$, consistent with the fact that such an increase drives the system towards phase separation.
Moreover, by looking at the large-times displacement distributions in Figs.~\ref{fig:vanhove-loc}(b) and \ref{fig:vanhove-loc}(d), we notice that it takes longer for the system at higher hard colloid density to balance out the two contributions. 
Overall, population splitting leads to an effective two-state dynamics for the hard colloids, which we rationalize by building a two-state toy model.

\subsection{Switching model for population dynamics}
\label{sec:results:switchingmodel}

At a mesoscopic length scale of the order of $r_c$, defined in Eq.~\eqref{eq:locdensity} 
employed to compute $\rho_\mathrm{loc}^i(t)$, hard colloids switch between two diffusive states, a fast and a slow one, corresponding to diffusion in rich-colloid regions and poor-colloid regions, respectively. 
We assume that each diffusive state is characterized by a different diffusion coefficient, $D_\mathrm{f}$ and $D_\mathrm{s}$, respectively (with $D_\mathrm{f}>D_\mathrm{s}$), 
The characteristic time to switch from the fast to the slow state is $\tau_\mathrm{f}$, while the time to switch from the slow to the fast state is $\tau_\mathrm{s}$, with $\tau_\mathrm{f}\le\tau_\mathrm{s}$. 
We then consider a one-dimensional model in which, at each time $t$ and coarse-grained position $x$, the total number of hard colloids is given by the sum of colloids that are in the fast and slow states. 
Defining the corresponding number density $n_{\mathrm{slow}(\mathrm{fast})}=N_{\mathrm{slow}(\mathrm{fast})}/N_H$, we have $n_\mathrm{TOT}(x,t)=n_\mathrm{slow}(x,t)+n_\mathrm{fast}(x,t)$, with $\int_{-\infty}^{\infty}n_\mathrm{TOT}(x,t) {\mathrm d}x =1$.
The hard colloid motion can be described using the following coupled diffusion equations,~\cite{doerries2022apparent}
\begin{eqnarray}
    \frac{\partial}{\partial t} n_\mathrm{slow}(x,t) &=& - \frac{n_\mathrm{slow}(x,t)}{\tau_\mathrm{s}} + \frac{n_\mathrm{fast}(x,t)}{\tau_\mathrm{f}} + D_\mathrm{s} \frac{\partial^2}{\partial x^2} n_\mathrm{slow}(x,t), \label{eq:n-slow}\\
    \frac{\partial}{\partial t} n_\mathrm{fast}(x,t) &=& - \frac{n_\mathrm{fast}(x,t)}{\tau_\mathrm{f}} + \frac{n_\mathrm{slow}(x,t)}{\tau_\mathrm{s}} + D_\mathrm{f} \frac{\partial^2}{\partial x^2} n_\mathrm{fast}(x,t), 
    \label{eq:n-fast}
\end{eqnarray}
given some initial conditions $n^0_{\mathrm{slow}}(x)\equiv n_\mathrm{slow}(x,t=0)$ and $n^0_{\mathrm{fast}}(x)\equiv n_\mathrm{fast}(x,t=0)$. 
As an alternative, one could also define a Langevin-like description with a switching diffusion coefficient that can be solved making use of the subordination technique defined in
Refs.~\onlinecite{carollo2024two,sposini2024being}.
We can solve our set of equations \eqref{eq:n-slow}-\eqref{eq:n-fast} in the Fourier--Laplace ($\{k,u\}$) space. 
In particular, if we denote with $\hat{\mathcal O}(k,t)$ the Fourier transform and with $\tilde{\mathcal O}(x,u)$ the Laplace transform of the quantity
${\mathcal O}(x,t)$, such that ${\hat {\tilde {\mathcal O}}}(k,u)$ 
is the combined Fourier--Laplace transform, 
we obtain:
\begin{eqnarray}
    {\hat {\tilde {n}}}_\mathrm{slow}(k,u) &=& {\hat n}_\mathrm{slow}^0(k) \frac{\tau_\mathrm{s} \tau_\mathrm{f} \left(u+1/\tau_\mathrm{f} + D_\mathrm{f} k^2 \right)}{\tau_\mathrm{s} \tau_\mathrm{f} \left(u+1/\tau_\mathrm{f} + D_\mathrm{f} k^2 \right)\left(u+1/\tau_\mathrm{s} + D_\mathrm{s} k^2 \right)-1} \nonumber \\
    & + & {\hat n}_\mathrm{fast}^0(k) \frac{\tau_\mathrm{s}}{\tau_\mathrm{s} \tau_\mathrm{f} \left(u+1/\tau_\mathrm{f} + D_\mathrm{f} k^2 \right)\left(u+1/\tau_\mathrm{s} + D_\mathrm{s} k^2 \right)-1}; \\
    {\hat {\tilde {n}}}_\mathrm{fast}(k,u) &=& {\hat n}_\mathrm{fast}^0(k) \frac{\tau_\mathrm{s} \tau_\mathrm{f} \left(u+1/\tau_\mathrm{s} + D_\mathrm{s} k^2 \right)}{\tau_\mathrm{s} \tau_\mathrm{f} \left(u+1/\tau_\mathrm{f} + D_\mathrm{f} k^2 \right)\left(u+1/\tau_\mathrm{s} + D_\mathrm{s} k^2 \right)-1}\nonumber \\
    & + & {\hat n}_\mathrm{slow}^0(k) \frac{\tau_\mathrm{f}}{\tau_\mathrm{s} \tau_\mathrm{f} \left(u+1/\tau_\mathrm{f} + D_\mathrm{f} k^2 \right)\left(u+1/\tau_\mathrm{s} + D_\mathrm{s} k^2 \right)-1},
\end{eqnarray}
from which we can easily obtain ${\hat {\tilde n}}_\mathrm{TOT}(k,u)={\hat {\tilde n}}_\mathrm{slow}(k,u)+{\hat {\tilde n}}_\mathrm{fast}(k,u)$.
Finally, we can use our solution in Fourier--Laplace space to calculate the moments in the Laplace space,
\begin{equation}
    \langle {\tilde x}^n(u)\rangle = (-{\mathrm i})^n \frac{\partial^n}{\partial k^n} {\hat {\tilde n}}_\mathrm{TOT}(k,u)\Big|_{k=0}.
    \label{eq:moments}
\end{equation}
We are interested in understanding how the interplay between the two populations affects the particle dynamics, particularly the long time dynamics. 
We thus focus on the simplest case in which all particles start at position $x=0$, with a fraction of colloids $f_\mathrm{s}$ starting in the slow state and another fraction $f_\mathrm{f} = 1 - f_\mathrm{s}$ starting in the fast state. 
This implies the initial conditions $n_\mathrm{slow}^0(x)  =  f_\mathrm{s} \delta(x)$ and $n_\mathrm{fast}^0(x)  =  f_\mathrm{f} \delta(x)$, with Fourier transforms ${\hat n}_\mathrm{slow}^0(k) = f_\mathrm{s}$ and ${\hat n}_\mathrm{fast}^0(k) = f_\mathrm{f}$, respectively.
In this way, by using Eq.~(\ref{eq:moments}) and computing the inverse Laplace transform, we obtain explicit expressions for the second and fourth moments:
\begin{eqnarray}
    \langle x^2(t)\rangle &=& 2 \, \left\{ \frac{D_\mathrm{s}\tau_\mathrm{s}+D_\mathrm{f}\tau_\mathrm{f}}{\tau_\mathrm{s}+\tau_\mathrm{f}}  \, t \label{eq:2mom} 
    + \frac{(D_\mathrm{s}-D_\mathrm{f}) \tau_\mathrm{f}\tau_\mathrm{s} (f_\mathrm{s} \tau_\mathrm{f}-f_\mathrm{f}\tau_\mathrm{s})}{(\tau_\mathrm{s}+\tau_\mathrm{f})^2}\left[1-
    \exp\left(-{t}/{\bar{\tau}} \right)\right] \right \}, \\
    \langle x^4(t)\rangle &=&  12 \, \frac{(D_\mathrm{s}\tau_\mathrm{s}+D_\mathrm{f}\tau_\mathrm{f})^2}{(\tau_\mathrm{s}+\tau_\mathrm{f})^2}  \, t^2 \nonumber \\
    & & + 24 \, \frac{(D_\mathrm{f}-D_\mathrm{s})^2 \tau_\mathrm{f}^2\tau_\mathrm{s}^2\left[f_\mathrm{s} \tau_\mathrm{f}(\tau_\mathrm{f}-2\tau_\mathrm{s})+ f_\mathrm{f} \tau_\mathrm{s}(\tau_\mathrm{s}-2\tau_\mathrm{f})\right]}{(\tau_\mathrm{s}+\tau_\mathrm{f})^4} 
    \left[1-\exp\left(-{t}/{\bar{\tau}} \right)\right]
    \nonumber
    \\
    & & + 24 \, \frac{(D_\mathrm{f}-D_\mathrm{s}) \tau_\mathrm{f}\tau_\mathrm{s}}{(\tau_\mathrm{s}+\tau_\mathrm{f})^3} \, t  \nonumber 
    \left\{ (D_\mathrm{s}\tau_\mathrm{f} + D_\mathrm{f}\tau_\mathrm{s})(f_\mathrm{s} \tau_\mathrm{f}-f_\mathrm{f} \tau_\mathrm{s}) \exp\left(-{t}/{\bar{\tau}} \right)  
    \right. 
    \nonumber \\
    & & \left.  
    - D_\mathrm{s}\tau_\mathrm{s} \left[2 f_\mathrm{s} \tau_\mathrm{f} + f_\mathrm{f}(\tau_\mathrm{f}-\tau_\mathrm{s})  \right]  
    - D_\mathrm{f}\tau_\mathrm{f} \left[f_\mathrm{s}\tau_\mathrm{f}-\tau_\mathrm{s}(2f_\mathrm{f}+f_\mathrm{s})\right] \right\}
    \label{eq:4mom}
\end{eqnarray}
where we defined $\bar{\tau}=\tau_\mathrm{s} \tau_\mathrm{f}/(\tau_\mathrm{s} +\tau_\mathrm{f})$.
From Eq.~(\ref{eq:2mom}) it can be seen that, for $t \gg \bar{\tau}$, our model 
shows a diffusive regime $\langle x^2(t)\rangle \simeq 2 D_{LT} t$, where we indicate with $D_{LT} =(D_\mathrm{s}\tau_\mathrm{s}+D_\mathrm{f}\tau_\mathrm{f})/(\tau_\mathrm{s}+\tau_\mathrm{f})$ the long-time diffusion coefficient.
Combining Eqs.~(\ref{eq:2mom}) and (\ref{eq:4mom}) we can calculate the NG parameter, which reads
\begin{equation}
\alpha_2(t)=\frac{\langle x^4(t)\rangle}
{3\, \langle x^2(t)\rangle^2}-1 \simeq C \left(t/\bar{\tau}\right)^{-1}, \quad t \gg \bar{\tau},
\label{eq:alpha2-toymodel}
\end{equation}
where 
\begin{equation}
C=\frac{2 (D_\mathrm{f}-D_\mathrm{s})}{(D_\mathrm{s}\tau_\mathrm{s}+D_\mathrm{f}\tau_\mathrm{f})^2} \left[(f_\mathrm{f} \tau_\mathrm{s} - f_\mathrm{s} \tau_\mathrm{f})(D_\mathrm{s}\tau_\mathrm{s} + D_\mathrm{f}\tau_\mathrm{f}) + \tau_\mathrm{s}\tau_\mathrm{f}(D_\mathrm{f}-D_\mathrm{s})\right].
\label{eq:coeffNG}
\end{equation}
Note that if $D_\mathrm{f} = D_\mathrm{s}$ the NG vanishes, as it should, because we no longer have two distinct populations. 
Our results show that the presence of different populations of diffusivity leads to the emergence of non Gaussian dynamics, even when the MSD effectively displays a simple diffusive trend; a phenomenon known as Brownian yet non Gaussian dynamics, largely discussed in literature in the last two decades.~\cite{wang2012brownian,metzler2020superstatistics,doerries2022apparent, sposini2024being}  
We highlight in particular the $t^{-1}$ scaling of the NG parameter at long times reported in Eq.~\eqref{eq:alpha2-toymodel}, which is in line with results obtained with other similar models,~\cite{hurley1996non,sposini2024being} and identifies the contribution from the population splitting to the overall non-Gaussian dynamics of the hard colloids.
We thus proceed in comparing our simulations with this analytical results.  
\begin{figure}
    \centering
    \xincludegraphics[width=0.49\linewidth,label=a)]{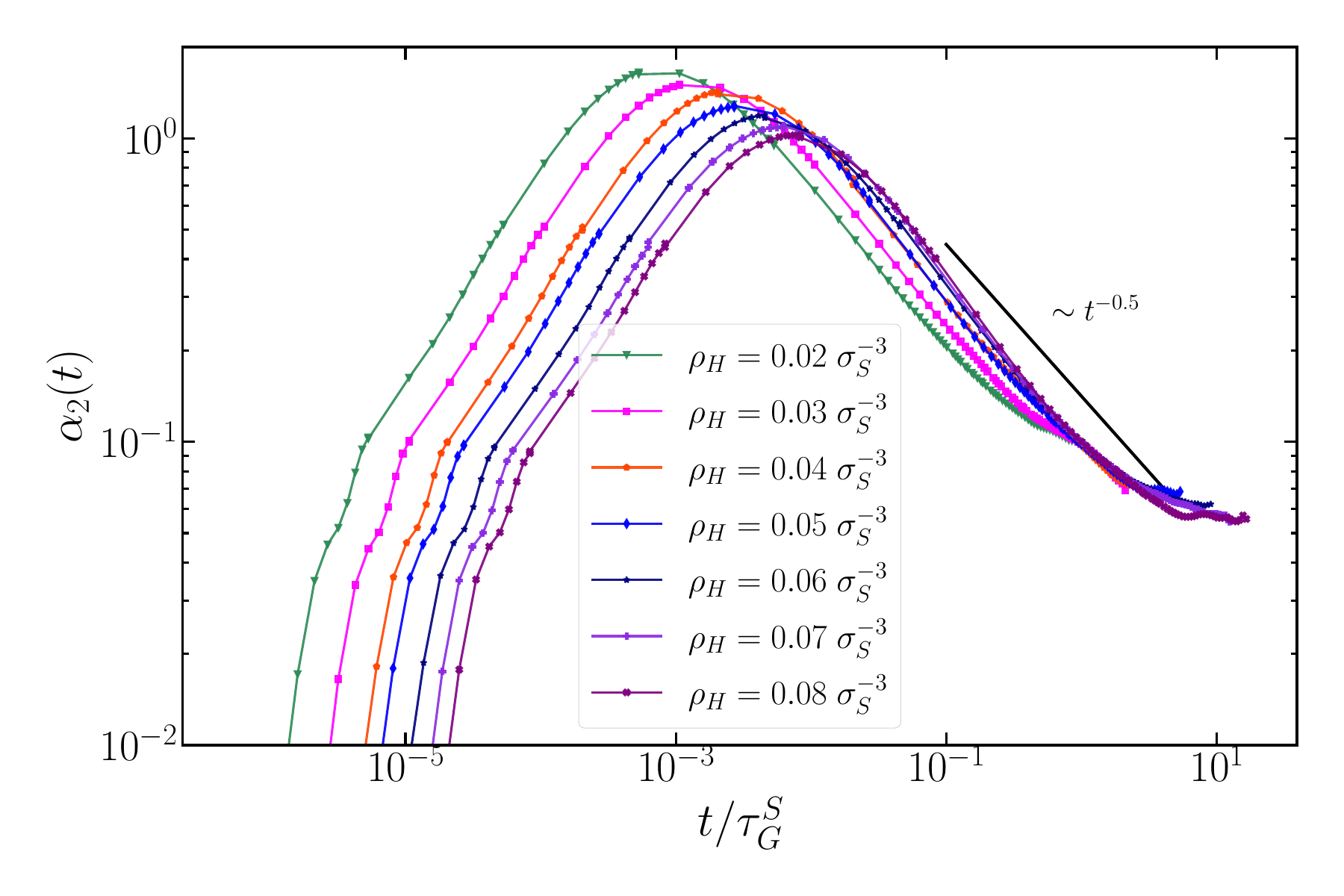} 
    \xincludegraphics[width=0.49\linewidth,label=b)]{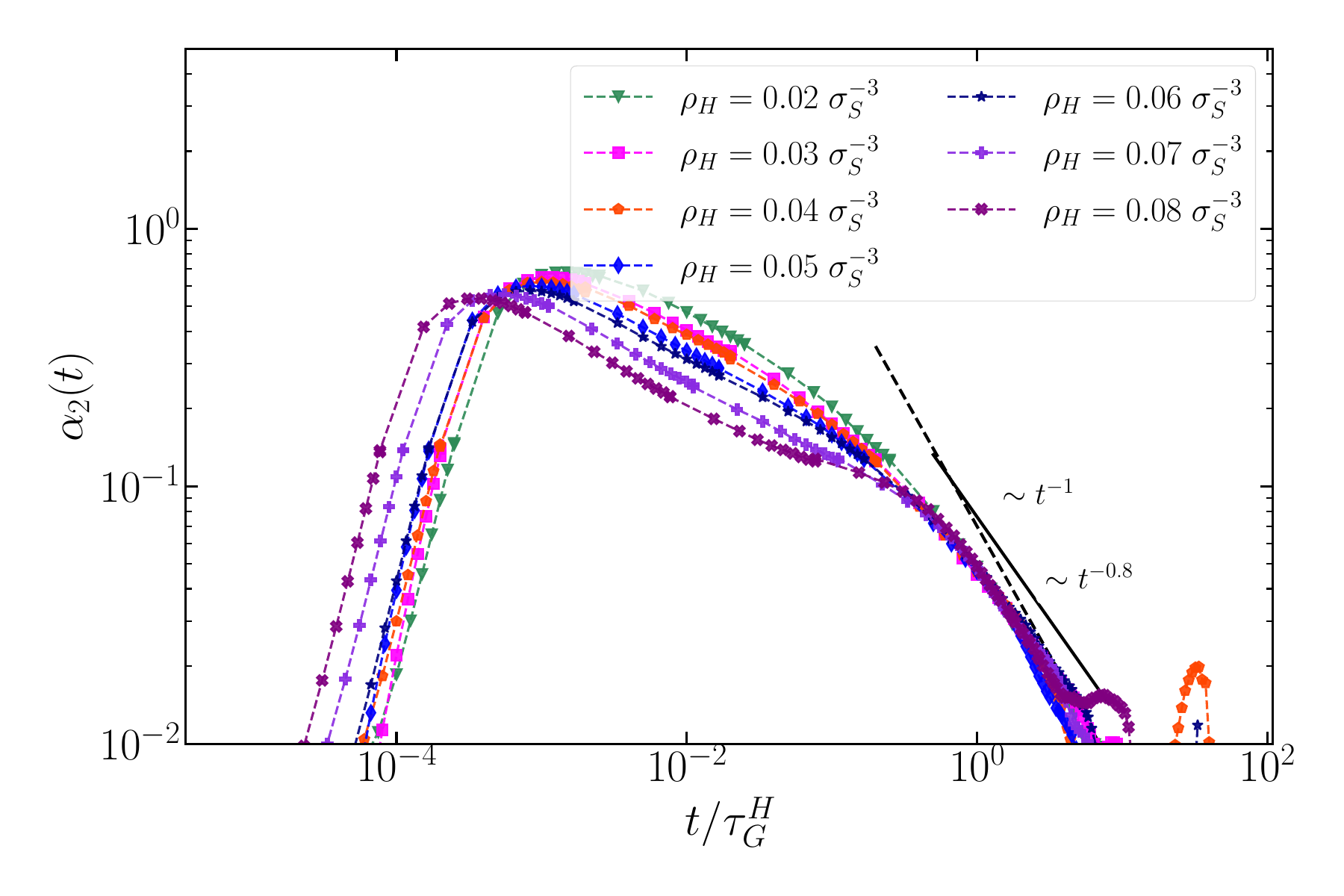} 
    \caption{Non-Gaussian parameter of soft (dashed lines) and hard (solid lines) components for different values of $\rho_H$, obtained from MD simulations. The time for the soft [(a)] and hard [(b)] component non-Gaussian parameters is rescaled by $\tau_G^{S}$, and  $\tau_G^{H}$, respectively.}
    \label{fig:nongauss}
\end{figure}
Following the idea developed by 
Rusciano {\em et al.} in Ref.~\onlinecite{rusciano2022fickian} we define the time-scale that identifies the recovery of Gaussianity as that time $\tau_G$ at which $\alpha_2(\tau_G)< \alpha_2^\star$, where $\alpha_2^\star$ is a low threshold value. 
The displacement distribution is indistinguishable from a Gaussian distribution for $t\ge \tau_G$.
We compute this time-scale both for the soft- and hard-component, 
$\tau_G^{S}$ (obtained with $\alpha_2^\star=0.1$) and $\tau_G^{H}$ (obtained with $\alpha_2^\star=0.05$), respectively, and use them to rescale the NG parameters, as shown in Fig.~\ref{fig:nongauss}.
In particular, we find that by rescaling the NG parameter of the soft component by $\tau_G^{S}$ we highlight the power-law decay $\sim t^{-\delta}$, with $\delta<1$, for all $\rho_H$ values---see Fig.~\ref{fig:nongauss}(a). 
For glassy systems it was indeed observed that, depending on the interparticle potential, one can observe different power law exponents, from $\delta=0.55$ for hard spheres to $\delta=0.7$ for soft disks.~\cite{rusciano2022fickian}
Note that the curves do not show a full collapse, as the threshold $\alpha_2^\star$ is not reached within the monitored time (this is why we used a value of $\alpha_2^\star$ for stars larger than the one used for hard colloids).  
The NG parameter of the hard component, instead, when rescaled by $\tau_G^{H}$, nicely shows a full collapse at long times, as reported in Fig.~\ref{fig:nongauss}(b). 
In particular, we observe a universal power law trend for all $\rho_H$ with an exponent $\delta\cong 0.8$.
This result is not fully in line with Eq.~\eqref{eq:alpha2-toymodel} obtained from the switching diffusion model, suggesting that both the glassiness of the matrix and the population splitting contribute to the trend of the NG parameter in a non trivial way. 

\section{Conclusions and outlook}
\label{sec:concl}
In this work, we have investigated the structural and dynamical behavior of a binary mixture composed of soft and hard colloidal components as a function of the hard colloid concentration. 
Our molecular dynamics simulations reveal a rich interplay between glassy dynamics, demixing, and emergent population splitting.
From the structural analysis, we observed that while the short- and intermediate-range correlations remain largely unaffected by variations in hard colloid density, both the soft–soft and hard–hard structure factors develop pronounced low-$q$ peaks with increasing $\rho_H$.
This behavior signals the onset of long-range correlations and an incipient phase separation. 
The concentration–concentration structure factor further confirms the emergence of a demixing process, displaying a growing and shifting secondary peak consistent with the formation of hard colloid–rich regions.
Notably, our simulations capture clear signatures of arrested phase separation already at relatively low $\rho_H$, preceding the full demixing expected from the equilibrium phase diagram.
The analysis on the dynamics of our system provides complementary insights. 
On the one hand, the mean-square displacements, the self van Hove distribution functions, and corresponding self-intermediate scattering functions indicate that increasing the hard colloid concentration progressively melts the soft glassy matrix. 
On the other hand, the hard colloids behave as diffusive tracers embedded in a heterogeneous medium, exhibiting a crossover from diffusive to logarithmic relaxation at intermediate time and length scales. 
The self van Hove functions and non-Gaussian parameters reveal strong dynamical heterogeneity, with the soft component displaying the characteristic signatures of glassy dynamics and the hard component developing an additional source of non-Gaussianity associated with population splitting induced by arrested phase separation.
By quantifying the local density distributions of the hard colloids, we established a direct link between spatial heterogeneity and dynamical heterogeneity: particles in colloid-rich regions are significantly more mobile than those in colloid-poor regions. 
This observation is rationalized through a two-state diffusion model that captures the alternating motion between high- and low-diffusivity environments. 
Furthermore, the collective dynamics of the hard component exhibits a pronounced decoupling from self-dynamics at low $q$, reinforcing the impact of structural heterogeneity on the collective relaxation mechanisms.
Overall, our results demonstrate that the interplay between glassy dynamics and arrested phase separation gives rise to complex, multi-scale behavior in soft–hard colloidal mixtures. 
The coexistence of different motility populations, coupled with the progressive melting of the soft glass, highlights the nontrivial coupling between structure, dynamics, and composition. 

Our findings can be verified in experiments on mixtures of soft star polymers an hard colloids, which are readily possible,~\cite{parisi:jcp:2021} where it would be useful to be able to track the small colloidal tracers only, in a fashion similar to what has already been done for binary hard sphere mixtures.~\cite{sentjabrskaja2016anomalous}
Although the predicted split of the colloidal into two populations, a fast and a small one, would be hard to track, the logarithmic decay of their incoherent relaxation function while at the same time the mean-square displacements would remain insensitive to the tracer concentration, would be a strong corroboration of the scenario put forward in this work. 
From the theoretical and computational perspective, future investigations should be directed towards the inclusion of enthalpic interactions between the two components and of active tracer particles, as well as towards improving the modeling of the populations dynamics of hard colloids. 
In particular, generalizations to a more space-resolved model could help to understand the deviations from the expected analytical results of the NG parameter at long times. 

\begin{acknowledgments}
We gratefully acknowledge Dr.~Francesco Arceri for useful suggestions and stimulating discussions and Prof.~Daniele Parisi for providing us with a copy of Figure~1.
K.~N.~M.~acknowledges support from the Erasmus+ programme of the European Union. 
V.~S.~acknowledges support from the European Union's Horizon Europe research and innovation programme under the Marie Sk{\l}odowska-Curie Actions (MSCA), grant agreement number 101149450.
\end{acknowledgments}

\section*{Data Availability Statement}
The data that support the findings of this study are available from the corresponding author upon reasonable request. 

\section*{Author Contributions}
{\bf Konstantin~N.~Moser:} Performing the simulations (lead); formal analysis (lead); methodology (supporting); software (lead); visualization (lead); writing - original draft (supporting). {\bf Christos N.~Likos:} Conceptualization (lead); formal analysis (supporting); methodology (equal); funding acquisition (lead); project administration (equal); resources (lead); supervision (equal); validation (equal); writing - original draft (supporting). {\bf Vittoria~Sposini:} Conceptualization (lead); formal analysis (equal); methodology (equal); funding acquisition (supporting); project administration (equal); resources (supporting); supervision (equal); validation (equal); writing - original draft (lead).

\bibliography{final-bib}

\end{document}